\documentclass[11pt]{article}
\usepackage[utf8]{inputenc}
\usepackage{textcomp}
\usepackage{amsfonts}
\usepackage{amsmath}
\usepackage{amssymb}
\usepackage{mathrsfs}
\usepackage{bbold}
\usepackage[margin=2.2cm]{geometry}
\usepackage{longtable}
\usepackage{lscape,graphicx}
\usepackage{graphics}
\usepackage{color}
\usepackage{array}
\usepackage{makecell}
\usepackage{bm}
\usepackage{comment}
\usepackage{multicol}
\usepackage{blindtext}
\usepackage{float}

\usepackage{hyperref}
\usepackage{cleveref}

\usepackage[numbers, sort&compress]{natbib}

\usepackage{slashed}

\usepackage{sidecap}
\usepackage{caption}
\usepackage{subcaption}

\newcommand{\ba}{\begin{array}{c}}
\newcommand{\ea}{\end{array}}

\newcommand{\be}{\beta}

\usepackage[dvipsnames]{xcolor}

\def\be{\begin{equation}}
\def\ee{\end{equation}}
\def\beq{\begin{equation}}
\def\eeq{\end{equation}}
\def\bc{\begin{center}}
\def\ec{\end{center}}
\def\bea{\begin{eqnarray}}
\def\eea{\end{eqnarray}}

\begin{document}
\begin{titlepage}
\vspace*{-1cm}
\phantom{hep-ph/***}
\flushright

\vskip 1.5cm
\begin{center}
\mathversion{bold}
{\LARGE\bf
Lepton mixing and charged lepton flavour violation from inverse seesaw with non-degenerate heavy states
\mathversion{normal}
}
\vskip .3cm
\end{center}
\vskip 0.5  cm
\begin{center}
{\large F.~P.~Di Meglio} and
{\large C.~Hagedorn}
\\
\vskip .7cm
{\footnotesize
Instituto de F\'isica Corpuscular, CSIC and Universidad de Valencia,
Edificio Institutos Investigaci\'on, Catedr\'atico Jos\'e Beltr\'an 2, 46980 Paterna, Spain
\vskip .5cm
\begin{minipage}[l]{.9\textwidth}
\begin{center}
\textit{E-mail:}
\tt{francescopaolo.dimeglio@ific.uv.es}, \tt{claudia.hagedorn@ific.uv.es}
\end{center}
\end{minipage}
}
\end{center}
\vskip 1cm
\begin{abstract}
We analyse an inverse seesaw scenario with 3+3 gauge singlets. The flavour structure is determined by a flavour symmetry, $\Delta (3 \, n^2)$ or $\Delta (6 \, n^2)$, $n$ integer, and CP 
and their residual groups among charged leptons and the neutral states. For the latter, the Dirac mass matrix of the gauge singlets 
carries all non-trivial flavour structure. Consequently, the heavy sterile states form three pseudo-Dirac pairs which have in general 
distinct masses. We discuss the signal strength of different charged lepton flavour
violating processes. Ensuring that the lepton mixing angles can be accommodated at the $3 \, \sigma$ level or better, we find that the current bounds on 
the branching ratios of $\mu\to e \gamma$, $\mu \to 3 \, e$, $\tau\to \ell \, \gamma$ and $\tau \to 3 \, \ell$, $\ell=e, \mu$, as well as the rate of $\mu-e$ conversion in nuclei do not strongly constrain
the considered parameter space, while the limits expected from the upcoming experiments Mu3E, COMET and Mu2e will have a relevant impact.
 \end{abstract}
\end{titlepage}
\setcounter{footnote}{0}

\section{Introduction}
\label{intro}

Neutrino masses and the observed lepton mixing pattern, encoded in the Pontecorvo-Maki-Nakagawa-Sakata (PMNS) mixing matrix, are two observational facts that cannot be satisfactorily explained in the Standard Model (SM). 
For this reason, many extensions of the SM have been proposed that feature a mechanism for neutrino mass generation. Among the most known ones are 
the type-I~\cite{typeIseesaw1,typeIseesaw2,typeIseesaw3,typeIseesaw4,typeIseesaw5}, type-II~\cite{typeIIseesaw1,typeIIseesaw2,typeIIseesaw3,typeIIseesaw4,typeIIseesaw5,typeIIseesaw6} 
and type-III seesaw mechanisms~\cite{typeIIIseesaw}. If the involved couplings are of order one and no (generalised) lepton number is imposed, see e.g.~\cite{Shaposhnikov:2006nn,Kersten:2007vk,Moffat:2017feq},
 the new particles are expected to have masses (much) larger than a few TeV. In contrast, the inverse seesaw (ISS) mechanism~\cite{Mohapatra:1986bd,GonzalezGarcia:1988rw,Mohapatra:1986aw,Bernabeu:1987gr}
 can explain the lightness of neutrinos with the help of particle masses in the reach of current (and near-future) experiments and sizeable couplings, since the smallness of neutrino masses 
 originates from the small breaking of lepton number. Non-abelian discrete symmetries, also in combination with CP, have shown to be prime candidates to understand the
 peculiar properties of lepton mixing, in particular if these symmetries are broken to (distinct) residual groups among charged leptons and neutral states~\cite{GfCP,GfCPothers1,GfCPothers2}, 
 see~\cite{GfCPearly1,GfCPearly2,GfCPearly3,GfCPearly4,GfCPearly5,GfCPearly6} for earlier works 
 and~\cite{King:2015aea,Feruglio:2019ybq,Ishimori:2010au,Gfreviewnew,Grimus:2011fk} for reviews. Members of the series of groups
 $\Delta (3 \, n^2)$~\cite{Luhn:2007uq} and $\Delta (6 \, n^2)$~\cite{Escobar:2008vc}, $n$ integer, are examples for such non-abelian discrete symmetries. As is known~\cite{Hagedorn:2014wha},
  from these, combined with CP, four different types of lepton mixing patterns can be derived that correspond to Case 1), Case 2), Case 3 a) and Case 3 b.1), respectively.
A concise description of their properties can be found in e.g.~\cite{Hagedorn:2021ldq}. For further studies of lepton mixing from the groups $\Delta (3 \, n^2)$
and $\Delta (6 \, n^2)$ and CP, 
see e.g.~\cite{DeltaCPothers1,DeltaCPothers2,DeltaCPothers3,DeltaCPothers4,DeltaCPothers5,DeltaCPothers6,DeltaCPothers7,DeltaCPothers8,DeltaCPothers9,DeltaCPothers10,DeltaCPothers11}.

In the current work, we consider an ISS mechanism with $3+3$ heavy sterile states, $N_i$ and $S_j$, $i,j=1,2,3$. 
As flavour group we use one of the groups $\Delta (3 \, n^2)$ and $\Delta (6 \, n^2)$ (together with an additional $Z_3$ symmetry $Z_3^{\mathrm{(aux)}}$),
while the CP symmetry corresponds to an automorphism of the flavour group. The charged lepton mass matrix is determined by a residual $Z_3$ group such that it is diagonal and contains three free parameters encoding the 
charged lepton masses. The neutrino Yukawa coupling and the mass matrices fixing the mass spectrum of the heavy sterile states are instead subject to the 
residual group $G_\nu=Z_2 \times CP$. Three different minimal cases can be studied: option 1 for which only the Majorana mass matrix of the sterile states $S_j$ breaks the original
symmetry to $G_\nu$~\cite{Hagedorn:2021ldq}, while for option 2 only the neutrino Yukawa coupling encodes the breaking to $G_\nu$~\cite{DiMeglio:2024gve} and option 3
is characterised by a Dirac mass matrix coupling the heavy sterile states $N_i$ and $S_j$ that is only invariant under $G_\nu$. Here, we analyse the phenomenology of the latter option, in particular
the masses and mixing of the light and heavy neutral states as well as the signal strength of charged lepton flavour violation (cLFV) such as the branching ratio (BR) of $\mu\to e \gamma$ and the $\mu-e$
conversion rate (CR) in nuclei.

The remainder of the paper is organised as follows: section~\ref{setup} contains an outline of the setup, in which
the flavour and CP symmetries and the transformation properties of the different fields under these are presented as well as 
the residual symmetries of the different mass matrices are discussed and their form is shown. Results for the mass spectrum
of the light neutrinos and the heavy sterile states can be found in section~\ref{massesmixing} together with the mixing
of light and heavy sterile states. In section~\ref{analytics} the masses of the heavy sterile states and the BRs and CR
of $\mu-e$ transitions are analytically investigated. Results of numerical scans for different examples of Case 1)
through Case 3 b.1) are discussed in section~\ref{numerics}. Furthermore, we briefly comment on the expected size of BRs of radiative and trilepton 
cLFV tau lepton decays. We compare the results obtained for option 3 with those for option 1, found in~\cite{Hagedorn:2021ldq},
and for option 2, see~\cite{DiMeglio:2024gve}, in section~\ref{comparison}. In section~\ref{summ} we summarise. 
 We collect supplementary plots in appendix~\ref{app:suppplots}.

\section{Setup}
\label{setup}

We employ as flavour symmetry $G_f$ a member of the series of groups $\Delta (3 \, n^2)$ and $\Delta (6 \, n^2)$, $n$ not divisible by three nor by four.
Left-handed (LH) lepton doublets $L_\alpha$, $\alpha=e, \mu, \tau$, as well as the singlets $N_i$ and $S_j$, $i,j,=1,2,3$, transform
as a triplet under $G_f$, which we specify further below. Right-handed (RH) charged leptons $\ell_{\alpha R}$ are instead singlets (all in ${\bf 1}$)
under this group. In order to distinguish these a $Z_3$ symmetry $Z_3^{\mathrm{(aux)}}$ is used, i.e.~$\ell_{e R} \sim 1$, $\ell_{\mu R} \sim \omega$ and  
$\ell_{\tau R} \sim \omega^2$ with $\omega= e^{\frac{2 \, \pi i}{3}}$. All other fields are invariant under $Z_3^{\mathrm{(aux)}}$.
The CP symmetry is chosen according to~\cite{GfCP,GfCPothers1,GfCPothers2}, see also~\cite{GfCPearly1,GfCPearly2,GfCPearly3,GfCPearly4,GfCPearly5,GfCPearly6}, as automorphism of $G_f$.

As mentioned, the charged lepton mass matrix $m_\ell$ is diagonal with three independent entries, corresponding to $m_e$, $m_\mu$ and $m_\tau$.\footnote{We always assume them to be ordered canonically.}
This can be achieved with a residual $Z_3$ symmetry that is the diagonal subgroup of the $Z_3$ symmetry, generated by the generator $a$ of the flavour group (which is always diagonal itself, compare e.g.~appendix A 
in~\cite{DiMeglio:2024gve}), and $Z_3^{\mathrm{(aux)}}$.

The Lagrangian relevant for the masses and the mixing of the neutral states reads as follows
\begin{equation}
- (y_D)_{\alpha i} \, \overline{L}^c_\alpha \, H \, N^c_i - (M_{NS})_{ij} \, \overline{N}_i \, S_j - \frac 12\, (\mu_S)_{kl} \, \overline{S}^c_k \, S_l + \mathrm{h.c.}
\end{equation}
with $\langle H \rangle \approx 174 \, \mbox{GeV}$, being the vacuum expectation value (VEV) of the SM Higgs. With the neutrino Yukawa coupling matrix $y_D$ we can define the Dirac mass matrix $m_D$, $m_D= y_D \, \langle H \rangle$.
The mass matrices, $m_D$, $M_{NS}$ and $\mu_S$, are constrained in general by the residual symmetry $G_\nu$. We are interested in studying minimal cases. Two of these, option 1 and 2,
in which either $\mu_S$ or $m_D$ are invariant under $G_\nu$, while the other mass matrices possess the original symmetry, have already been scrutinised in the literature~\cite{Hagedorn:2021ldq,DiMeglio:2024gve}.
For the remaining option, option 3, the form of the matrix $M_{NS}$ is non-trivial in flavour space, while $m_D$ and $\mu_S$ preserve the entire flavour group and CP.
This option can be realised as follows: we assign the singlets $S_j$ to the real, unfaithful irreducible representation ${\bf 3^\prime}$ such that $\mu_S$ non-zero is allowed in the limit
of unbroken symmetries and its flavour structure is trivial,\footnote{This form of $\mu_S$ arises, since complex representation matrices for ${\bf 3^\prime}$ are employed.}
\begin{equation}
\label{eq:formmuS}
\mu_S = \mu_0 \, 
\left(
\begin{array}{ccc}
1 & 0 & 0\\
0 & 0 & 1\\
0 & 1 & 0
\end{array}
\right) \;\;\; \mbox{with} \;\;\; \mu_0 > 0 \; .
\end{equation}
LH lepton doublets $L_\alpha$ are assigned to a complex, faithful irreducible representation of $G_f$, called ${\bf 3}$, such that
lepton mixing is determined by the structure of $G_f$ and its residual symmetries. As the Dirac mass matrix $m_D$ should also be invariant under the original group, we assign the singlets $N_i$
to ${\bf 3}$ as well. Then, the form of $m_D$ is -- in the limit of unbroken $G_f$ and CP -- given by
\begin{equation}
\label{eq:formmD}
m_D = y_0 \, 
\left(
\begin{array}{ccc}
1 & 0 & 0\\
0 & 1 & 0\\
0 & 0 & 1
\end{array}
\right) \; \langle H \rangle \;\;\; \mbox{with} \;\;\;  y_0 > 0 \; .
\end{equation}
Since $N_i \sim {\bf 3}$ and $S_j \sim {\bf 3^\prime}$, it is obvious that the matrix $M_{NS}$ vanishes for unbroken $G_f$. It is, thus, the source of symmetry breaking among the neutral states.
Its flavour structure is derived by requesting it to be invariant under the subgroup $G_\nu = Z_2 \times CP$. We denote as $Z ({\bf r})$
and $X ({\bf r})$ the generator of the $Z_2$ group and the CP transformation in the representation ${\bf r}$ (which equals ${\bf 3}$ for $L_\alpha$ and $N_i$ and 
${\bf 3^\prime}$ for $S_j$), respectively. Consequently, the conditions that $M_{NS}$ has to fulfil are
\begin{equation}
Z ({\bf 3})^\dagger \, M_{NS} \, Z ({\bf 3^\prime}) = M_{NS} \;\;\; \mbox{and} \;\;\; X ({\bf 3})^\star \, M_{NS} \, X ({\bf 3^\prime}) = M_{NS}^\star \; .
\end{equation}
The explicit form of the matrices $Z ({\bf 3})$, $Z ({\bf 3^\prime})$, $X ({\bf 3})$ and $X ({\bf 3^\prime})$ can be found in~\cite{Drewes:2022kap}. The form of $M_{NS}$ is constrained to be 
\begin{equation}
\label{eq:formMNS}
M_{NS} = \Omega ({\bf 3}) \, R_{ij} (\theta_N) \, \mathrm{diag} (M_1, M_2, M_3) \, P^{ij}_{kl} \, R_{kl} (-\theta_S) \, \Omega ({\bf 3^\prime})^\dagger 
\end{equation}
with the unitary matrices $\Omega ({\bf 3})$ and $\Omega ({\bf 3^\prime})$ obeying
\begin{equation}
\Omega ({\bf 3}) \, \Omega ({\bf 3})^T = X ({\bf 3}) \;\;\; \mbox{and} \;\;\; \Omega ({\bf 3^\prime}) \, \Omega ({\bf 3^\prime})^T = X ({\bf 3^\prime}) \; .
\end{equation}
Furthermore, $R_{ij} (\theta_N)$ and $R_{kl} (-\theta_S)$ are two rotations in the $(ij)$- and $(kl)$-planes, fixed by the pair of degenerate eigenvalues 
of the matrices $Z ({\bf 3})$ and $Z ({\bf 3^\prime})$ in the basis transformed by $\Omega ({\bf 3})$ and $\Omega ({\bf 3^\prime})$, respectively. The permutation matrix $P^{ij}_{kl}$
is only necessary, if these two planes do not coincide. The two angles $\theta_N$ and $\theta_S$ together with the three mass parameters $M_1$, $M_2$ and $M_3$ are 
the five (real) parameters contained in the matrix $M_{NS}$. For each case, Case 1) through Case 3 b.1), we have different forms of $\Omega ({\bf 3})$ and $\Omega ({\bf 3^\prime})$
as well as the rotation planes $(ij)$ and $(kl)$, see~\cite{Drewes:2022kap}.\footnote{We note that the form of $M_{NS}$ is exactly the same as the one of the neutrino Yukawa coupling matrix
in~\cite{Drewes:2022kap}.} As can be seen in the next section, the three mass parameters $M_1$, $M_2$ and $M_3$ correspond to the distinct masses of the three pseudo-Dirac pairs of heavy neutral states to a very high degree.

\section{Masses and mixing of light and heavy sterile states}
\label{massesmixing}

In the following, we study the masses and mixing of the neutral states arising from the diagonalisation, 
\begin{equation}
{\cal U}^T \, {\cal M}_{\mathrm{Maj}} \, {\cal U} = \mathrm{diag} \left( m_1, m_2, m_3, m_4, \dots, m_9 \right)   \; ,
\end{equation}
of the nine-by-nine matrix ${\cal M}_{\mathrm{Maj}}$,
\begin{equation}
\label{eq:MMaj}
{\cal M}_{\mathrm{Maj}} = \left(
\begin{array}{ccc}
\mathbb{0} & m_D & \mathbb{0} \\
m_D^T & \mathbb{0} & M_{NS} \\
\mathbb{0} & M_{NS}^T & \mu_S
\end{array}
\right) \; ,
\end{equation}
written in the basis $(\nu_{\alpha L}, N_i^c, S_j)$, with the unitary nine-by-nine matrix ${\cal U}$,
\begin{equation}
\label{eq:Ucal}
{\cal U} = \left(
\begin{array}{cc}
\tilde{U}_\nu & S \\
T & V
\end{array}
\right) \; .
\end{equation}
The masses $m_{1,2,3}$ are those of the light neutrinos and $m_{4,...,9}$ of the six heavy sterile states, while $\tilde{U}_\nu$ is a three-by-three, $S$ a three-by-six, $T$ a six-by-three and $V$ a six-by-six matrix.
None of these are in general unitary. 

With the form of $\mu_S$, $m_D$ and $M_{NS}$, see eqs.~(\ref{eq:formmuS},\ref{eq:formmD},\ref{eq:formMNS}), and assuming that $|\mu_S| \ll |m_D| \ll |M_{NS}|$~\cite{Mohapatra:1986bd,GonzalezGarcia:1988rw,Mohapatra:1986aw,Bernabeu:1987gr}, 
we first deduce the mass matrix of the heavy sterile states and their mass spectrum. We have to look at
\begin{equation}
\left( 
\begin{array}{cc}
\mathbb{0} & M_{NS}\\
 M_{NS}^T& \mu_S
 \end{array}
 \right)
\end{equation}
which is approximately diagonalised by the matrix $V$, i.e.~
\begin{equation}
V^T\, \left( 
\begin{array}{cc}
\mathbb{0} & M_{NS}\\
 M_{NS}^T& \mu_S
 \end{array}
 \right) \, V \approx \mathrm{diag} \left( m_4, \dots, m_9 \right) \; .
\end{equation}
We get
\begin{equation}
\label{eq:formV}
V = \frac{1}{\sqrt{2}} \, \left(
\begin{array}{cc}
i \, \Omega ({\bf 3})^\star \, R_{ij} (\theta_N) & \Omega ({\bf 3})^\star \, R_{ij} (\theta_N)\\
-i \, \Omega ({\bf 3^\prime}) \, R_{kl} (\theta_S) \, (P^{ij}_{kl})^T & \Omega ({\bf 3^\prime}) \, R_{kl} (\theta_S) \, (P^{ij}_{kl})^T 
\end{array}
\right) \; .
\end{equation}
The masses of the heavy sterile states, that form pseudo-Dirac pairs, are given by
\begin{equation}
\label{eq:heavymasses}
m_4 \approx m_7 \approx M_1 \;\; , \; m_5 \approx m_8 \approx M_2 \;\;\; \mbox{and} \;\;\; m_6 \approx m_9 \approx M_3  \; .
\end{equation}
These are splitted among each other by terms of order $\mu_0$. Indeed, if the combination 
\begin{equation}
\label{eq:comb}
P^{ij}_{kl} \, R_{kl} (-\theta_S) \, \Omega ({\bf 3^\prime})^T \, \left( \begin{array}{ccc}
1 & 0 & 0 \\
0 & 0 & 1 \\
0 & 1 & 0
\end{array}
\right) \, \Omega ({\bf 3^\prime}) \, R_{kl} (\theta_S) \, (P^{ij}_{kl})^T
\end{equation}
is diagonal, the masses of the heavy sterile states read
\begin{equation}
\label{eq:m49Mfdiag}
\!\!\!\!\!\!
m_4 \approx M_1 - \frac{\mu_0}{2} \;\; , \; m_7 \approx M_1 + \frac{\mu_0}{2} \;\; , \;
m_5 \approx M_2 - \frac{\mu_0}{2} \;\; , \; m_8 \approx M_2 + \frac{\mu_0}{2} \;\; , \;
m_6 \approx M_3 - \frac{\mu_0}{2} \;\; , \; m_9 \approx M_3 + \frac{\mu_0}{2}  \; .
\end{equation}
If the combination in eq.~(\ref{eq:comb}) is not diagonal, but only block-diagonal, then only the masses of one pair of pseudo-Dirac particles
can be written in this form. For e.g.~Case 1), it is
\begin{equation}
m_5 \approx M_2 - \frac{\mu_0}{2} \;\; , \; m_8 \approx M_2 + \frac{\mu_0}{2} \; ,
\end{equation}
while the other masses are -- up to the first order in $\mu_0$ --
\begin{equation}
\label{eq:m49MfnondiagCase1}
\!\!\!\!\!\!
m_4 \approx M_1 - \frac{\mu_0}{2} \, \cos 2 \, \theta_S \;\; , \; m_7 \approx M_1 + \frac{\mu_0}{2} \, \cos 2 \, \theta_S \;\; , \; 
m_6 \approx M_3 + \frac{\mu_0}{2} \, \cos 2 \, \theta_S \;\; , \; m_9 \approx M_3 - \frac{\mu_0}{2} \, \cos 2 \, \theta_S \; .
\end{equation}
Coming to the leading order contribution to the light neutrino mass matrix $m_\nu$~\cite{Hettmansperger:2011bt}, 
\begin{equation}
\label{eq:mnuLO}
m_\nu = m_D \, \Big( M_{NS}^{-1} \Big)^T \, \mu_S \, M_{NS}^{-1} \, m_D^T \; ,
\end{equation}
we see that it is of the form
\begin{eqnarray}
\label{eq:mnuform}
\nonumber
m_\nu&=& y_0^2 \, \mu_0 \, \langle H \rangle^2 \, U_0 (\theta_N)^\star \, \mathrm{diag} (M_1^{-1}, M_2^{-1}, M_3^{-1}) 
\\
&&\nonumber \;\, \left[ P^{ij}_{kl} \, R_{kl} (-\theta_S) \, \Omega ({\bf 3^\prime})^T \, \left(
\begin{array}{ccc}
1 & 0 & 0\\
0 & 0 & 1\\
0 & 1 & 0
\end{array}
\right) \, \Omega ({\bf 3^\prime}) \, R_{kl} (\theta_S) \, (P^{ij}_{kl})^T  \right] 
\\
&& \;\;\; \, \mathrm{diag} (M_1^{-1}, M_2^{-1}, M_3^{-1}) \, U_0 (\theta_N)^\dagger \; ,
\end{eqnarray}
where we have used the definition
\begin{equation}
\label{eq:U0def}
U_0 (\theta) = \Omega ({\bf 3}) \, R_{ij} (\theta) \; .
\end{equation}
Note that the expression in square brackets coincides with the one found in eq.~(\ref{eq:comb}) and it is decisive whether it is diagonal or block-diagonal.
If it is diagonal, we arrive at 
\begin{equation}
m_\nu =  y_0^2 \, \mu_0 \, \langle H \rangle^2 \, U_0 (\theta_N)^\star \, \mathrm{diag} (M_1^{-2}, M_2^{-2}, M_3^{-2}) \, U_0 (\theta_N)^\dagger \; ,
\end{equation}
meaning that we can determine the mixing matrix $\tilde{U}_\nu$,
\begin{equation}
\tilde{U}_\nu^T \, m_\nu \tilde{U}_\nu \approx \mathrm{diag} (m_1, m_2, m_3) \; ,
\end{equation}
to be of the form
\begin{equation}
\label{eq:UnutcasediagLO}
\tilde{U}_\nu \approx U_0 (\theta_N) = \Omega ({\bf 3}) \, R_{ij} (\theta_N) 
\end{equation}
and the light neutrino masses $m_f$, $f=1,2,3$, are equal to
\begin{equation}
\label{eq:mfMf}
m_f = \frac{y_0^2 \, \mu_0 \, \langle H \rangle^2}{M_f^2} \; .
\end{equation}
From eq.~(\ref{eq:UnutcasediagLO}) and taking into account that the charged lepton mass matrix is diagonal with canonically ordered masses, meaning $U_\ell = \mathbb{1}$,
 the (non-unitary) PMNS mixing matrix reads the same as $\tilde{U}_\nu$ and the lepton mixing parameters only depend on the free parameter $\theta_N$. 
 
We remark that an additional permutation is necessary for Case 3 b.1), that requires a different assignment of the light neutrino masses, 
\begin{equation}
 \label{eq:mfMf_Case3b1}
m_1 = \frac{y_0^2 \, \mu_0 \, \langle H \rangle^2}{M_3^2} \; , \;\;
m_2 = \frac{y_0^2 \, \mu_0 \, \langle H \rangle^2}{M_1^2}  \; , \;\; 
m_3 = \frac{y_0^2 \, \mu_0 \, \langle H \rangle^2}{M_2^2} \; ,
\end{equation}
cf.~\cite{Hagedorn:2014wha,Drewes:2022kap,DiMeglio:2024gve}.
For simplicity, we focus on Case 1) through Case 3 a) in the following, where this additional permutation is not needed for the lepton mixing pattern. However, all formulae can be easily also derived for Case 3 b.1) 
and all statements made hold analogously.
 
If the expression in eq.~(\ref{eq:comb}) is not diagonal, an additional rotation whose angle is fixed by two of the mass parameters $M_f$ and the free angle $\theta_S$ is required, and it occurs in the same
plane as $R_{ij} (\theta_N)$. So, we replace $\theta_N$ by an effective angle $\bar{\theta}_N$. 
 At the same time, the light neutrino masses $m_f$ with $f=i$ or $f=j$ are not directly proportional to $M_f^{-2}$ anymore, while the remaining mass still is.

The sub-leading contribution to the light neutrino mass matrix, $m_\nu^1$, in the expansion in powers of $(|m_D|/|M_{NS}|)$~\cite{Hettmansperger:2011bt}, reads in general
\begin{equation}
\!\!\!\!\!\!\!\!\!m_\nu^1 = -\frac 12 \, m_D \, \Big( M_{NS}^{-1} \Big)^T \, \left( \mu_S \, M_{NS}^{-1} \, m_D^T \, m_D^\star \, \Big( M_{NS}^{-1} \Big)^\dagger + \Big( M_{NS}^{-1} \Big)^\star \, m_D^\dagger \, m_D \, \Big( M_{NS}^{-1} \Big)^T \, \mu_S \right) \, M_{NS}^{-1} \, m_D^T \; .
\end{equation}
It is convenient to rewrite $m_\nu^1$ with the help of the (hermitean) matrix $\eta$ which encodes the non-unitarity of the matrix $\tilde{U}_\nu$ (and, thus, of the PMNS mixing matrix), induced by the mixing of the light and heavy sterile states. It is defined as
\begin{equation}
\eta = \frac 12 \, m_D^\star \, (M_{NS}^{-1})^\dagger \, M_{NS}^{-1} \, m_D^T \; , 
\end{equation}
and, consequently, $m_\nu^1$ takes the form
\begin{equation}
m_\nu^1= - \left( m_\nu \, \eta + \eta^\star \, m_\nu \right) \; .
\end{equation}
Using $m_D$ and $M_{NS}$ as in eqs.~(\ref{eq:formmD},\ref{eq:formMNS}), the matrix $\eta$ is given by
\begin{equation}
\label{eq:etadiag}
\eta = \eta_0^{\prime\prime} \, U_0 (\theta_N) \, \mbox{diag} (M_1^{-2}, M_2^{-2}, M_3^{-2}) \, U_0 (\theta_N)^\dagger \; ,
\end{equation}
where
\begin{equation}
\label{eq:eta0pp}
\eta_0^{\prime\prime} = \frac{y_0^2 \, \langle H \rangle^2}{2}\; .
\end{equation}
So, the matrix $m_\nu^1$ reads 
\begin{equation}
m_\nu^1 = - 2 \, \eta_0^{\prime\prime} \, \tilde{U}_\nu^\star \, \mbox{diag} \left(\frac{m_1}{M_1^2}, \frac{m_2}{M_2^2}, \frac{m_3}{M_3^2}\right) \, \tilde{U}_\nu^\dagger \; ,
\end{equation}
if the expression in eq.~(\ref{eq:comb}) is diagonal. In this situation, the sub-leading contribution does not affect the lepton mixing parameters and only slightly impacts the light neutrino masses.

If instead the combination in eq.~(\ref{eq:comb}) is non-diagonal, the matrix $U_0 (\theta_N)$ appearing in $\eta$ cannot be identified with $\tilde{U}_\nu$, because $\tilde{U}_\nu \approx U_0 (\bar{\theta}_N)$.
However, we can still write $\eta$ in terms of $\tilde{U}_\nu$ as
\begin{equation}
\eta = \eta_0^{\prime\prime} \, \tilde{U}_\nu \, R_{ij} (\theta_N - \bar{\theta}_N) \, \mbox{diag} (M_1^{-2},M_2^{-2},M_3^{-2}) \,  R_{ij} (\theta_N - \bar{\theta}_N)^T \,  \tilde{U}_\nu^\dagger \; . 
\end{equation}
Then, we find
\begin{equation}
m_\nu^1 = - \eta_0^{\prime\prime} \, \tilde{U}_\nu^\star \, [ \mbox{diag} (m_1,m_2,m_3) \, R_{ij} (\theta_N - \bar{\theta}_N) \, \mbox{diag} (M_1^{-2},M_2^{-2},M_3^{-2}) \, R_{ij} (\theta_N - \bar{\theta}_N)^T + 
\mbox{transpose} ] \, \tilde{U}_\nu^\dagger \, , 
\end{equation}
meaning that this contribution introduces a small additional rotation in the $(ij)$-plane as well as small shifts in the light neutrino masses.

The (non-unitary) PMNS mixing matrix is of the form
\begin{equation}
\label{eq:UPMNSetaU0}
\tilde{U}_{\mathrm{PMNS}}= \tilde{U}_\nu = (\mathbb{1}-\eta) \, U_0 \; .
\end{equation}
For the expression in eq.~(\ref{eq:comb}) being diagonal, the matrix $U_0$ equals $U_0 = U_0 (\theta_N)$ and using $\eta$ as in eq.~(\ref{eq:etadiag}) we have 
\begin{equation}
\label{eq:nonuni_diag}
\tilde{U}_{\mathrm{PMNS}} = U_0 (\theta_N) \, (\mathbb{1}- \eta_0^{\prime\prime} \, \mbox{diag} (M_1^{-2},M_2^{-2},M_3^{-2})) \; .
\end{equation}
We, thus, expect that the columns of the PMNS mixing matrix receive different suppression factors, since $\eta^{\prime\prime}_0$ is positive and the mass parameters $M_f$ are in general different.
Indeed, the latter can be expressed in terms of the light neutrino masses, see eq.~(\ref{eq:mfMf}). 

If the expression in eq.~(\ref{eq:comb}) is not diagonal, the matrix $U_0$ in eq.~(\ref{eq:UPMNSetaU0}) is $U_0 = U_0 (\bar{\theta}_N)$ and, thus, we get
\begin{equation}
\label{eq:nonuni_nondiag}
\tilde{U}_{\mathrm{PMNS}} = U_0 (\bar{\theta}_N) \, (\mathbb{1}- \eta_0^{\prime\prime} \, R_{ij} (\theta_N-\bar{\theta}_N) \, \mbox{diag} (M_1^{-2},M_2^{-2},M_3^{-2}) \, R_{ij} (\theta_N-\bar{\theta}_N)^T) \; .
\end{equation}
Again, the correction acts differently on the different columns of the PMNS mixing matrix. 

Eventually, we mention the form of $S$ and $T$, that are computed -- at leading order -- as
\begin{equation}
S = \left(
\begin{array}{ccc}
\mathbb{0} &,& m_D^\star \, (M_{NS}^{-1})^\dagger  
\end{array}
\right) \, V 
\;\;\; \mbox{and} \;\;\;
T = \left( 
\begin{array}{c}
\mathbb{0} \\ - M_{NS}^{-1} \, m_D^T \, \tilde{U}_\nu
\end{array}
\right) \; .
\end{equation}
While the expression of $S$ is always
\begin{equation}
\label{eq:formS}
S = \frac{y_0 \, \langle H \rangle}{\sqrt{2}} \, \left(
\begin{array}{ccc}
-i \, U_0 (\theta_N) \, \mbox{diag} (M_1^{-1},M_2^{-1},M_3^{-1}) &,& U_0 (\theta_N) \, \mbox{diag} (M_1^{-1},M_2^{-1},M_3^{-1})
\end{array}
\right) \; ,
\end{equation}
the form of the matrix $T$ instead depends on whether the combination in eq.~(\ref{eq:comb}) is diagonal or not. For the diagonal case, we have
\begin{equation}
T = - y_0 \, \langle H \rangle \, \left( 
\begin{array}{c}
\mathbb{0} \\ \Omega ({\bf 3^\prime}) \, R_{kl} (\theta_S) \, (P^{ij}_{kl})^T \, \mbox{diag} (M_1^{-1},M_2^{-1},M_3^{-1}) 
\end{array}
\right) \; ,
\end{equation}
whereas for the non-diagonal case we find
\begin{equation}
T = -y_0 \, \langle H \rangle \, \left( 
\begin{array}{c}
\mathbb{0} \\ \Omega ({\bf 3^\prime}) \, R_{kl} (\theta_S) \, (P^{ij}_{kl})^T \, \mbox{diag} (M_1^{-1},M_2^{-1},M_3^{-1}) \, R_{ij} (\theta_N-\bar{\theta}_N)^T 
\end{array}
\right) \; .
\end{equation}

\section{Analytical estimates}
\label{analytics}

In the following, we first estimate the expected size of the parameters $y_0$, $\mu_0$ and $M_f$ of this scenario 
and then the signal strength of certain cLFV processes.

\mathversion{bold}
\subsection{Estimates of parameters}
\mathversion{normal}
\label{analyticsMf}

\begin{figure}[t!]
    \centering
     \includegraphics[width=0.9\textwidth]{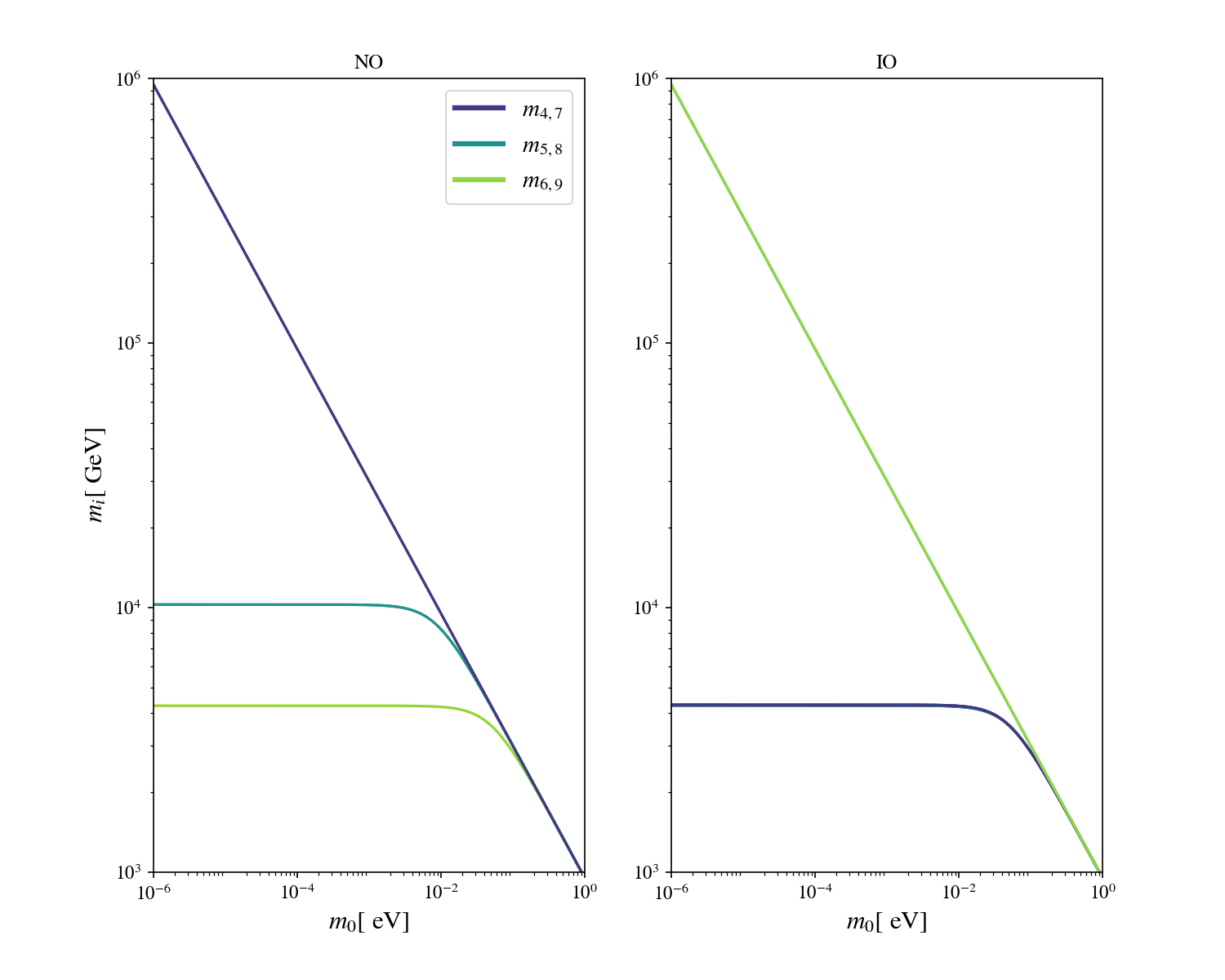}
    \caption{{\bf \mathversion{bold}Masses $m_{4,...,9}$ of the heavy sterile states in GeV as function of the lightest neutrino mass $m_0$ in eV\mathversion{normal}.} In the left plot we display the results for light neutrino masses with NO 
   and in the right plot for IO light neutrino masses. For concreteness, we fix $y_0=0.1$ and $\mu_0=3 \, \mathrm{keV}$ and use the best-fit values of the solar and the atmospheric mass squared differences from~\cite{Esteban:2024eli}.
   Note that the four masses $m_{4}$, $m_{5}$, $m_{7}$ and $m_{8}$ are almost degenerate in the case of an IO light neutrino mass spectrum, compare also eq.~(\ref{eq:miratiosIOm0large}).
    }
    \label{fig:Mfm0NOIO}
\end{figure}

To get an idea of the typical size of the Yukawa coupling $y_0$ and the scale $\mu_0$ of lepton number violation and the relations among the mass parameters $M_f$, $f=1,2,3$,
we express these in terms of the light neutrino masses. We focus on the situation in which the combination in eq.~(\ref{eq:comb}) is diagonal and assume one of the cases Case 1)
through Case 3 a) to be realised.  From eq.~(\ref{eq:mfMf}), we have for light neutrino masses with normal ordering (NO) 
\begin{equation}
\label{eq:Mfdiag}
M_1 = y_0 \, \langle H \rangle \, \sqrt{\frac{\mu_0}{m_0}} \;\; , \;\; M_2 = y_0 \, \langle H \rangle \, \sqrt{\frac{\mu_0}{\sqrt{m_0^2 + \Delta m_{21}^2}}} \;\; , \;\; M_3 = y_0 \, \langle H \rangle \, \sqrt{\frac{\mu_0}{\sqrt{m_0^2 + \Delta m_{31}^2}}} \;\; ,
\end{equation}
where $m_0$ is the lightest neutrino mass, while $\Delta m_{21}^2$ and $\Delta m_{31}^2$ are the solar and the atmospheric mass squared differences, respectively. 
We use the experimental best-fit values of the latter from~\cite{Esteban:2024eli} and take $m_0=0.03 \, \mbox{eV}$, which is the maximal value of the lightest neutrino mass compatible with cosmological bounds on the sum of neutrino masses, see~\cite{Planck:2018vyg}. For mass parameters in the interval $150 \, \mbox{GeV}$ to $10 \, \mbox{TeV}$, we estimate the size of $y_0$ and $\mu_0$ to be
\begin{equation}
0.2 \lesssim y_0 \, \sqrt{\frac{\mu_0}{\mbox{eV}}} \lesssim 10 \; .
\end{equation}
Using the information on the light neutrino mass spectrum, we can also estimate the ratios of the mass parameters to be 
\begin{equation}
\frac{M_1}{M_3} \approx 1.40 \;\;\; \mbox{and}\;\;\; \frac{M_2}{M_3} \approx 1.37 \; .
\end{equation}
For the lightest neutrino mass being zero, we see that $M_1$ becomes large, while the ratio of $M_2$ and $M_3$ tends to $2.41$.
 These values agree well with the numerical results, see section~\ref{numerics}. 
 We display the dependence of the masses of the heavy sterile states on $m_0$ for fixed values of $y_0$ and $\mu_0$, $y_0=0.1$ and $\mu_0=3 \, \mbox{keV}$, in the left of 
fig.~\ref{fig:Mfm0NOIO}.\footnote{The shown plots are produced by solving numerically for the eigenvalues and eigenvectors of the nine-by-nine matrix ${\cal M}_{\mathrm{Maj}}$ for a concrete choice of case, group theory
parameters and CP symmetry that leads to the expression in eq.~(\ref{eq:comb}) being diagonal.}

Similarly, we have for light neutrino masses with inverted ordering (IO)
\begin{equation}
M_1 = y_0 \, \langle H \rangle \, \sqrt{\frac{\mu_0}{\sqrt{m_0^2 + |\Delta m_{32}^2| - \Delta m_{21}^2}}} \;\; , \;\; M_2 = y_0 \, \langle H \rangle \, \sqrt{\frac{\mu_0}{\sqrt{m_0^2 + |\Delta m_{32}^2|}}} \;\; , \;\; M_3 = y_0 \, \langle H \rangle \, \sqrt{\frac{\mu_0}{m_0}} \;\; ,
\end{equation}
with $\Delta m_{32}^2$ denoting the atmospheric mass squared difference. We can perform an analogous estimate, employing $m_0= 0.015 \, \mbox{eV}$ as maximal value compatible with cosmological 
observations, and find
\begin{equation}
0.2 \lesssim y_0 \, \sqrt{\frac{\mu_0}{\mbox{eV}}} \lesssim 7 \; .
\end{equation}
Using this value for $m_0$ and the experimental best-fit values of the mass squared differences, we arrive at the ratios
\begin{equation}
\label{eq:miratiosIOm0large}
\frac{M_1}{M_2} \approx 1.01 \;\;\; \mbox{and}\;\;\; \frac{M_3}{M_2} \approx 1.87 \; .
\end{equation} 
In the limit $m_0$ being zero, $M_3$ becomes large, whereas the ratio of $M_1$ and $M_2$ remains approximately $1.01$. 
The numbers are in agreement with the numerical results, found in section~\ref{numerics}.
The behaviour of the masses of the heavy sterile states with $m_0$ for fixed $y_0$ and $\mu_0$ ($y_0=0.1$ and $\mu_0=3 \, \mbox{keV}$) is presented in the right of fig.~\ref{fig:Mfm0NOIO}.

\mathversion{bold}
\subsection{Estimates for cLFV signals}
\mathversion{normal}
\label{analyticscLFV}

For these estimates we make use of the formulae found in~\cite{Alonso:2012ji,Ilakovac:1994kj}. As the heavy sterile states are in general not all (nearly) degenerate in mass, see preceding subsection,
we cannot apply some of the simplifications employed for option 1 and 2.

\begin{figure}[t!]
    \centering
     \includegraphics[width=\textwidth]{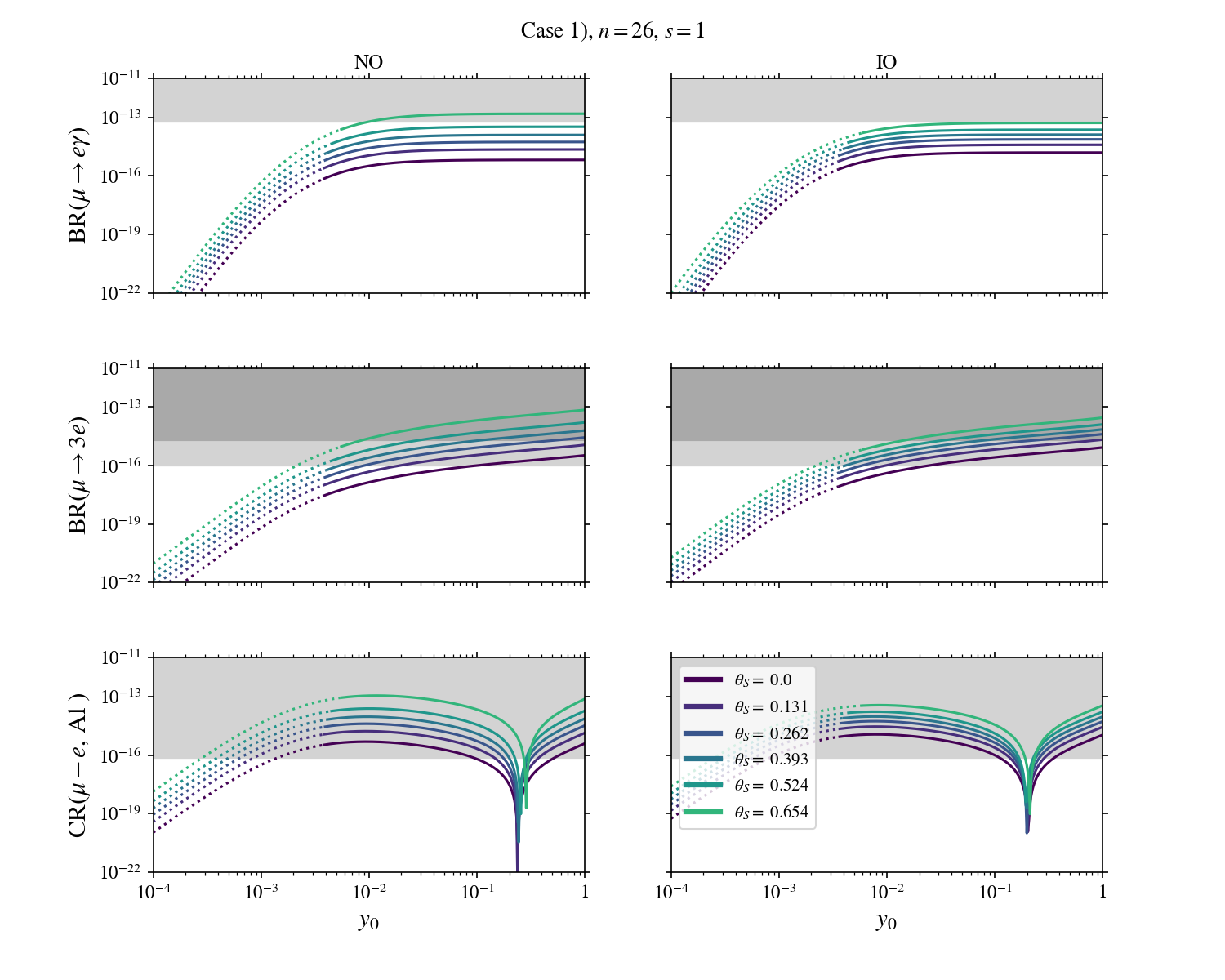}
    \caption{{\bf \mathversion{bold}Case 1). Results for $\mathrm{BR} (\mu\to e \, \gamma)$, $\mathrm{BR} (\mu\to 3 \,e)$ and $\mathrm{CR} (\mu-e, \mathrm{Al})$\mathversion{normal}} in the upper, middle and lower row, respectively. 
  These results are for $n=26$ and $s=1$, and six different values of the angle $\theta_S$, shown in different colours. In the left plot we display the results for light neutrino masses with NO and $m_0=0.03 \, \mathrm{eV}$
   and in the right plot for IO light neutrino masses with $m_0=0.015 \, \mathrm{eV}$. 
   For concreteness, we fix $\mu_0=3 \, \mathrm{keV}$ and use the best-fit values of the solar and the atmospheric mass squared differences and the reactor mixing angle~\cite{Esteban:2024eli}. The dotted lines
   indicate that at least one of the heavy sterile states has a mass lower than $150 \, \mathrm{GeV}$.
    }
    \label{fig:Case1mu0fixedNOIO}
\end{figure}

We first consider the radiative cLFV decays $\ell_\beta \to \ell_\alpha \, \gamma$ whose BR is given by 
\begin{equation}
\mathrm{BR} (\ell_\beta \to \ell_\alpha \, \gamma) = \frac{\alpha_w^3 \, s_w^2}{256 \, \pi^2} \, \frac{m_\beta^4}{M_W^4} \, \frac{m_\beta}{\Gamma_\beta} \, \left| G_\gamma^{\beta\alpha} \right|^2  \, ,
\end{equation} 
where $\alpha_w=\frac{g_w^2}{4 \, \pi}$ is the weak coupling, $s_w$ the sine of the weak mixing angle, $M_W$ the mass of the $W$ boson, $m_\beta$ the mass of the charged lepton of flavour $\beta$ and $\Gamma_\beta$ 
the corresponding total decay width. The form factor $G_\gamma^{\beta\alpha}$ reads
\begin{equation}
G_\gamma^{\beta\alpha} = \sum_{i=1}^{3+n_s} \, {\cal U}_{\alpha i} \, {\cal U}_{\beta i}^\star \, G_\gamma (x_i) \; ,
\end{equation}
where $n_s$ is the number of the sterile states ($n_s=6$), ${\cal U}$ the unitary matrix defined in eq.~(\ref{eq:Ucal}), $x_i=\left( \frac{m_i}{M_W} \right)^2$ and $G_\gamma (x)$ is the loop function 
\begin{equation}
G_\gamma (x) = -\frac{x \, (2 \, x^2 + 5 \, x -1)}{4 \, (1-x)^3} - \frac{3 \, x^3}{2 \, (1-x)^4} \, \mathrm{log} (x)
\end{equation}
with $G_\gamma (0)=0$ and the limit $G_\gamma (x) \approx \frac 12$ for $x \gg 1$. The contribution due to light neutrinos is negligible. Furthermore, we can use that the heavy sterile states form three
pairs of pseudo-Dirac neutrinos, compare eq.~(\ref{eq:heavymasses}), as well as the form of the matrix $S$ in eq.~(\ref{eq:formS}). In this way,
we can write the form factor $G_\gamma^{\beta\alpha}$ as
\begin{equation}
\label{eq:Ggba_approx}
G_\gamma^{\beta\alpha} \approx 2 \, \eta_0^{\prime\prime} \, \Big( U_0 (\theta_N)_{\alpha 2} \, U_0 (\theta_N)^\star_{\beta 2} \, \left( \frac{G_\gamma (x_5)}{M_2^2} - \frac{G_\gamma (x_4)}{M_1^2} \right)  
+ U_0 (\theta_N)_{\alpha 3} \, U_0 (\theta_N)^\star_{\beta 3} \, \left( \frac{G_\gamma (x_6)}{M_3^2} - \frac{G_\gamma (x_4)}{M_1^2} \right)  \Big)
\end{equation}
taking into account that $\alpha \neq \beta$. For larger masses of the heavy sterile states, i.e.~larger than $3.3 \, \mathrm{TeV}$,\footnote{This also holds for the numerical example in eq.~(\ref{eq:BRmuegestimate}).}
 the relative deviation of the loop function $G_\gamma (x)$ from $1/2$ is less than one 
percent and we can safely approximate $G_\gamma (x)$ as $1/2$ such that the expression for the form factor $G_\gamma^{\beta\alpha}$ simply reads
\begin{equation}
G_\gamma^{\beta\alpha} \approx \eta_{\alpha\beta} \; .
\end{equation}
Consequently, the BR of $\ell_\beta \to \ell_\alpha \, \gamma$ is proportional to $|\eta_{\alpha\beta}|^2$, as has also been found for option 2~\cite{DiMeglio:2024gve}.
These formulae can be used in order to estimate the size of the BR of $\mu\to e \gamma$ for each case. For Case 1) and assuming $\theta_S=0$,
 such that the matrix combination in eq.~(\ref{eq:comb}) is diagonal and the angle $\theta_N$ can be written in terms of the reactor mixing angle, cf.~\cite{Hagedorn:2014wha,Hagedorn:2021ldq}, we obtain
 \begin{equation}
 G_\gamma^{\mu e} \approx 2 \, \eta_0^{\prime\prime} \, \Big( \frac{1}{3} \, \left( \frac{G_\gamma (x_5)}{M_2^2} - \frac{G_\gamma (x_4)}{M_1^2} \right)  
- \frac 12 \, \sin \theta_{13}\, \left( \sin\theta_{13} + \sqrt{2-3 \, \sin^2 \theta_{13}}\right) \, \left( \frac{G_\gamma (x_6)}{M_3^2} - \frac{G_\gamma (x_4)}{M_1^2} \right)  \Big) \; .
 \end{equation}
 We note that the expression does not depend on the parameter $s$ that characterises the CP symmetry. 
 Using $y_0=0.1$, $\mu_0=3 \, \mathrm{keV}$ and the best-fit values of the two mass squared differences and the reactor mixing angle, the BR of the decay $\mu\to e \, \gamma$ is estimated as
 \begin{equation}
 \label{eq:BRmuegestimate}
 \mathrm{BR} (\mu\to e \, \gamma) \approx \; 6.6 \times 10^{-16} \; (1.6 \times 10^{-15}) 
 \end{equation} 
 for light neutrino masses with NO (IO) and $m_0=0.03 \, \mathrm{eV}$ ($m_0=0.015 \, \mathrm{eV}$). 
  Fig.~\ref{fig:Case1mu0fixedNOIO} shows for Case 1) ($n=26$ and $s=1$), different values of the angle $\theta_S$ and light neutrino masses with NO and IO the results for the BRs of $\mu\to e \, \gamma$ and $\mu\to 3 \, e$
 as well as the CR of $\mu-e$ conversion in aluminium for fixed $\mu_0$, $\mu_0=3 \, \mathrm{keV}$. The grey-shaded areas indicate the expected reach of the experiments MEG II, Mu3E (Phase 1 and Phase 2) and COMET, respectively   (see also section~\ref{prerequisites}). The values in eq.~(\ref{eq:BRmuegestimate}) agree well with those displayed in fig.~\ref{fig:Case1mu0fixedNOIO}.
 Similar estimates can be obtained for the other cases. Also the BRs of the trilepton cLFV decays $\ell_\beta \to 3 \, \ell_\alpha$ can be analysed in an analogous way.

 More interesting is the consideration of the CR of $\mu-e$ conversion, since we can investigate whether or not it is still possible to obtain an exact cancellation 
 of the CR for certain values of the parameters, in particular the value of the Yukawa coupling $y_0$. Indeed, the CR can be written in the following form
 \begin{equation}
 \label{eq:CRform}
 \mathrm{CR} (\mu-e, \mathrm{N}) \propto \left|U_0 (\theta_N)_{e2} \, U_0 (\theta_N)^\star_{\mu 2} \, \left( \frac{f (x_5)}{M_2^2} - \frac{f (x_4)}{M_1^2} \right)  
+ U_0 (\theta_N)_{e3} \, U_0 (\theta_N)^\star_{\mu 3} \, \left( \frac{f (x_6)}{M_3^2} - \frac{f (x_4)}{M_1^2} \right)\right|^2 \; ,
 \end{equation}
very similar to the expression for the BR of $\ell_\beta \to \ell_\alpha \, \gamma$, cf.~eq.~(\ref{eq:Ggba_approx}). The real-valued function $f(x)$ contains not only the nuclear 
form factors $D$, $V^{(n)}$ and $V^{(p)}$ of the nucleus $\mathrm{N}$, $s_w^2$ and the electric charge $e$, but also loop functions. Here, it is only relevant that in the limit $x\gg 1$, $f(x)$ goes to zero for
\begin{equation}
\label{eq:x0option2}
x_0 = \mathrm{exp} \left( \frac{\frac 98 \, V^{(n)}+ \left(\frac 98 +\frac{37 \, s_w^2}{12} \right) \, V^{(p)} - \frac{s_w^2}{16 \, e} \, D}{\frac 38 \, V^{(n)}+ \left( \frac{4 \, s_w^2}{3} -\frac 38\right) \, V^{(p)}}\right) \; ,
\end{equation}
compare~\cite{Alonso:2012ji}. The CR can vanish, if both the real and imaginary part of the expression in eq.~(\ref{eq:CRform}) are zero. In case that neither the real nor the imaginary part of 
$U_0 (\theta_N)_{e3} \, U_0 (\theta_N)^\star_{\mu 3}$ vanish, the relation
\begin{equation}
\label{eq:reimrel}
\rho = \frac{\mathrm{Re} (U_0 (\theta_N)_{e2} \, U_0 (\theta_N)^\star_{\mu 2})}{\mathrm{Re} (U_0 (\theta_N)_{e3} \, U_0 (\theta_N)^\star_{\mu 3})} 
=  \frac{\mathrm{Im} (U_0 (\theta_N)_{e2} \, U_0 (\theta_N)^\star_{\mu 2})}{\mathrm{Im} (U_0 (\theta_N)_{e3} \, U_0 (\theta_N)^\star_{\mu 3})}
\end{equation}
has to be fulfilled. Furthermore, assuming that the combination in eq.~(\ref{eq:comb}) is diagonal (either in general or because of a specific choice of the angle $\theta_S$), we can deduce from the vanishing of the real part alone 
\begin{equation}
\label{eq:x4Case13a}
x_4 = \mathrm{exp} \left( \frac{\frac 98 \, V^{(n)}+ \left(\frac 98 +\frac{37 \, s_w^2}{12} \right) \, V^{(p)} - \frac{s_w^2}{16 \, e} \, D}{\frac 38 \, V^{(n)}+ \left( \frac{4 \, s_w^2}{3} -\frac 38\right) \, V^{(p)}} 
+ \frac{\rho \, \frac{m_2}{m_1} \, \mathrm{log} \left(\frac{m_1}{m_2}\right) + \frac{m_3}{m_1} \, \mathrm{log} \left( \frac{m_1}{m_3} \right)}{1+\rho - \rho \, \left(\frac{m_2}{m_1} \right)- \left(\frac{m_3}{m_1} \right)} \right)
\end{equation}
for Case 1) through Case 3 a). For Case 3 b.1) instead, we have 
\begin{equation}
\label{eq:x4Case3b1}
x_4 = \mathrm{exp} \left( \frac{\frac 98 \, V^{(n)}+ \left(\frac 98 +\frac{37 \, s_w^2}{12} \right) \, V^{(p)} - \frac{s_w^2}{16 \, e} \, D}{\frac 38 \, V^{(n)}+ \left( \frac{4 \, s_w^2}{3} -\frac 38\right) \, V^{(p)}} 
+ \frac{\frac{m_1}{m_2} \, \mathrm{log} \left(\frac{m_2}{m_1}\right) + \rho \,\frac{m_3}{m_2} \, \mathrm{log} \left( \frac{m_2}{m_3} \right)}{1+\rho- \left(\frac{m_1}{m_2} \right) - \rho \, \left(\frac{m_3}{m_2} \right)} \right)
\end{equation}
due to the different relation between the light neutrino masses $m_f$ and the three mass parameters, $M_1$, $M_2$ and $M_3$, compare eqs.~(\ref{eq:mfMf}) and (\ref{eq:mfMf_Case3b1}). 
 Eqs.~(\ref{eq:x4Case13a}) and (\ref{eq:x4Case3b1}) clearly show that the value of $x_4$ at which the cancellation occurs is shifted with respect to the result obtained for  
option 2, see eq.~(\ref{eq:x0option2}). In addition, we observe that the value of $x_4$ depends on the case considered (through $\rho$ and the assignment of the light neutrino masses $m_f$ and the three mass parameters).
Expressing $m_f$ in eqs.~(\ref{eq:x4Case13a}) and (\ref{eq:x4Case3b1}) in terms of the lightest neutrino mass $m_0$ and the two measured mass squared differences also reveals that the value of $x_4$ changes with these and, obviously, 
 depends on the light neutrino mass ordering. In the limit in which $m_0$ is very small, e.g.~$m_0 \lesssim 10^{-5} \, \mathrm{eV}$, the form of the expressions in eqs.~(\ref{eq:x4Case13a}) and~(\ref{eq:x4Case3b1}) simplifies further.
For Case 1) and using the best-fit values of the two mass squared differences and the reactor mixing angle and employing $s_w^2=0.23$, $\alpha_{\mathrm{em}}=\frac{e^2}{4 \, \pi}=\frac{1}{137}$ as well as
the nuclear form factors for aluminium as found in~\cite{Kitano:2002mt}, we obtain from eq.~(\ref{eq:x4Case13a}) 
\begin{equation}
x_4 \approx 26829 \, (11033)
\end{equation} 
for light neutrino masses with NO (IO) and $m_0=0.03 \, (0.015) \, \mathrm{eV}$. Using again eq.~(\ref{eq:mfMf}) and setting $\mu_0= 3 \, \mathrm{keV}$, we can derive the corresponding value for $y_0$
\begin{equation}
y_0 \approx 0.24 \, (0.20) \; .
\end{equation}
 These values match well with the plots in the lower row of fig.~\ref{fig:Case1mu0fixedNOIO}. At least for light neutrino masses with NO a slight dependence on the value of the angle $\theta_S$
can be observed. Note that for Case 1) the imaginary parts of $U_0 (\theta_N)_{e2} \, U_0 (\theta_N)^\star_{\mu 2}$ and $U_0 (\theta_N)_{e3} \, U_0 (\theta_N)^\star_{\mu 3}$ are zero and, hence, 
the imaginary part of the expression in eq.~(\ref{eq:CRform}) always vanishes. Depending on the case and also on the value of the lightest neutrino mass $m_0$, in particular if $m_0$ is small, the value of 
$y_0$ is affected. Likewise, the choice of $\theta_S$ as well as the different values of $\theta_N$, that lead to a fit of the lepton mixing angles within the experimentally preferred $3 \,\sigma$ ranges~\cite{Esteban:2024eli},   can impact the value of $y_0$ at which the cancellation occurs (e.g.~for Case 3 a)). 
We also remark that even if an exact cancellation cannot happen, we encounter a minimum value of the CR for $\mu-e$ conversion in aluminium, e.g.~for
Case 2) compare figs.~\ref{fig:Case2mu0fixedNOIO} and~\ref{fig:Case2mu0fixedsmallNOIO} in appendix~\ref{app:suppplots}. Furthermore, we refer to the different figures showing results of the numerical scans, 
see section~\ref{numerics}.
However, in these the difference between an exact cancellation and a minimum in the CR for $\mu-e$ conversion in aluminium is difficult to observe.

Lastly, we comment on the possibility that a cancellation also occurs in (one of) the BRs. This could happen, since the masses of the heavy sterile states are in general not degenerate and the 
expressions for the BRs become more contrived. Nevertheless, such a cancellation is usually not observed in figs.~\ref{fig:Case1mu0fixedNOIO},\ref{fig:Case2mu0fixedNOIO},\ref{fig:Case2mu0fixedsmallNOIO} nor in the numerical scans, 
see section~\ref{numerics}, as it typically requires values of the Yukawa coupling $y_0$ (compare eq.~(\ref{eq:y0range})) and/or values of the masses of the heavy sterile states (see below eq.~(\ref{eq:m0benchmarks})) 
that are not considered in this study, and/or values of the angle $\theta_N$ which do not lead to an acceptable fit of the experimental data on lepton mixing. 
Some suppression in the BRs of $\mu\to e \gamma$ and $\mu\to 3 \, e$ can, however, be inferred in certain instances, compare e.g.~fig.~\ref{fig:Case1n26s1IOm010em5} in appendix~\ref{app:suppplots}.
Similar statements hold for the BRs of the cLFV tau lepton decays, $\tau\to e \gamma$, $\tau\to 3 \, e$, $\tau\to\mu \gamma$ and $\tau\to 3 \, \mu$, see section~\ref{taucLFVs}. 

\section{Numerical results}
\label{numerics}

We first enumerate the experimental inputs and (current/future) constraints as well as the region of parameter space of the scenario studied numerically. 
Then, we discuss results for the most relevant $\mu-e$ transitions, the decays $\mu\to e \gamma$ and $\mu\to 3 \, e$ as well as $\mu-e$ conversion in aluminium, 
for representatives of each case, Case 1) through Case 3 b.1). We also briefly comment on the cLFV tau lepton decays, $\tau\to e \gamma$, $\tau\to 3 \, e$,
$\tau\to\mu \gamma$ and $\tau\to 3 \, \mu$.

\subsection{Prerequisites}
\label{prerequisites}

We summarise the experimental inputs and constraints that we apply together with the region of parameter space scrutinised numerically.

\noindent {\bf Experimental inputs and constraints.} 
We employ the results of the global fit of the NuFIT collaboration for the best-fit values of the mass squared differences and the $3 \, \sigma$ ranges of 
the lepton mixing angles~\cite{Esteban:2024eli}. Furthermore, we use the bound on the sum of the light neutrino masses from cosmology~\cite{Planck:2018vyg}.
For the masses of the heavy sterile states, we require these to be at least $150 \, \mbox{GeV}$, compare~\cite{delAguila:2008cj,delAguila:2008hw,Chen:2011hc,Das:2012ze,Abada:2014vea,Arganda:2014dta,Abada:2014nwa,Abada:2014kba,Abada:2014cca,Arganda:2015naa,Abada:2015oba,DeRomeri:2016gum,Antusch:2016ejd,Crivellin:2022cve,Abada:2024hpb} for different phenomenological studies.

We use as bounds on the elements $\eta_{\alpha\beta}$, $\alpha,\beta=e,\mu,\tau$, those from~\cite{Blennow:2023mqx}.
 The current limits on the $\mu-e$ transitions, i.e.~the BR of $\mu\to e\gamma$ and $\mu\to 3 \, e$ as well as
 the CR of $\mu-e$ conversion in gold and titanium, are taken from~\cite{MEGII:2025gzr,SINDRUM:1987nra,SINDRUMII:2006dvw,SINDRUMIITi}, respectively.
As future bounds, we employ the one from MEG II for $\mu\to e \gamma$, $\mathrm{BR} (\mu \to e \gamma) < 6 \times 10^{-14}$~\cite{MEGII:2021fah}, 
from Mu3E for $\mu\to 3 \, e$, $\mathrm{BR} (\mu \to 3 \, e) < 20 \, (1) \times 10^{-16}$ Phase 1 (2)~\cite{Blondel:2013ia}, as well as 
from COMET (Mu2e) for $\mu-e$ conversion in aluminium, $\mathrm{CR} (\mu-e, \mathrm{Al}) < 7 \, (8) \times 10^{-17}$~\cite{COMET:2018auw,Jansen:2023ojv,Mu2e:2014fns,Artuso:2022ouk}, respectively.

For the BRs of the four cLFV tau lepton decays that we study we use the current limits from Belle~\cite{Belle:2021ysv,ParticleDataGroup:2020ssz}, Belle II~\cite{Belle-II:2024sce} and LHCb~\cite{LHCb:2026eod}. 
Prospective bounds on these BRs from Belle II, between a few times $10^{-10}$ and a few times $10^{-9}$, can be found in~\cite{Banerjee:2022xuw}.

\noindent {\bf Region of parameter space considered.} 
The parameter space of the neutral states comprises the Yukawa coupling $y_0$, the mass scale $\mu_0$ of lepton number violation, the three mass parameters $M_f$, $f=1,2,3$, as well as the two angles $\theta_N$ and $\theta_S$. 
We consider as range of $y_0$
\begin{equation}
\label{eq:y0range}
10^{-4} \leq y_0 \leq 1 \; ,
\end{equation}
while the scale $\mu_0$ is varied in the interval\footnote{This range is slightly extended compared to the numerical scan performed for option 2~\cite{DiMeglio:2024gve}.}
\begin{equation}
\label{eq:mu0range}
100 \, \mbox{eV} \leq \mu_0 \leq 1 \, \mbox{MeV} \; .
\end{equation}
Both of these quantities are varied logarithmically in their intervals.
We note that viable data points are found for $y_0 \gtrsim 2 \times 10^{-4}$, with the exact value slightly depending on the case considered, since only then the masses of the heavy
sterile states are larger than $150 \, \mbox{GeV}$ (and also all employed experimental constraints are satisfied). Likewise, the scale $\mu_0$ is constrained to fulfil $\mu_0 \gtrsim 2 \, \mbox{keV}$ (with the exact value again 
depending on the case considered) in order to pass 
current limits on the magnitudes of the elements of the matrix $\eta$ and the prospective bounds on the BRs of $\mu\to e \gamma$ and $\mu\to 3 \, e$ as well as on $\mathrm{CR} (\mu-e, \mathrm{Al})$.
 These findings are reflected in figs.~\ref{fig:Case1n26s1NOm0003}-\ref{fig:Case3an34s2m2NOm010em5_taudecays}.  
Furthermore, the unconstrained angle $\theta_S$, contained in $M_{NS}$, cf.~eq.~(\ref{eq:formMNS}), can lie in its entire admitted range
\begin{equation}
\label{eq:thetaSrange}
0 \leq \theta_S \leq 2 \, \pi \; ,
\end{equation}
which is scanned linearly in the numerical analysis.
In contrast, the angle $\theta_N$ is fixed in such a way that the measured lepton mixing angles are accommodated best (and always in their experimentally preferred $3 \, \sigma$ ranges)~\cite{Esteban:2024eli} and
 $M_f$, $f=1,2,3$, are adjusted in order to correctly reproduce the assumed light neutrino mass spectrum,
its ordering (NO or IO), the best-fit values of the solar and the atmospheric mass squared differences and the chosen benchmark
value of the lightest neutrino mass. The latter corresponds either to the maximal value compatible with cosmological observations~\cite{Planck:2018vyg},
\begin{equation}
\label{eq:m0benchmarks}
m_0 = 0.03 \, \mbox{eV} \;\; \mbox{(NO)} \;\; \mbox{or} \;\; m_0 = 0.015 \, \mbox{eV} \;\; \mbox{(IO)}, 
\end{equation}
or a benchmark value among the following ones, $m_0= 10^{-3} \, \mbox{eV}$, $m_0= 10^{-5} \, \mbox{eV}$, $m_0= 10^{-8} \, \mbox{eV}$, $m_0= 10^{-11} \, \mbox{eV}$ and $m_0= 10^{-14} \, \mbox{eV}$ -- for both light neutrino masses with NO and IO.
We remark, however, that for $m_0 \lesssim 10^{-5} \, \mbox{eV}$ the resulting plots look identical. We, thus, choose the value $m_0 =10^{-5} \, \mbox{eV}$.  
In this way, we typically get masses for the heavy sterile states ranging between $150 \, \mbox{GeV}$ and
$700 \, \mbox{TeV}$, with the upper value increasing with smaller values of $m_0$.

For certain plots, see e.g.~fig.~\ref{fig:Mfm0NOIO}, and numerical estimates, compare eq.~(\ref{eq:BRmuegestimate}) for example, we use as benchmark $y_0=0.1$ and $\mu_0=3 \, \mbox{keV}$.

The numerical scans are performed in an analogous way as in the study of option 2, cf.~details in appendix C of~\cite{DiMeglio:2024gve}.

\mathversion{bold}
\subsection{Scans for $\mu-e$ transitions}
\mathversion{normal}
\label{muetransitions}

We discuss representative examples for each case in turn and present corresponding plots. For all shown examples the lepton mixing angles are
accommodated at least at the $3 \, \sigma$ level~\cite{Esteban:2024eli}. In order to facilitate the comparison between the results obtained for option 3
and those for option 2~\cite{DiMeglio:2024gve}, we employ the same examples.

\begin{figure}[t!]
    \centering
     \includegraphics[width=\textwidth]{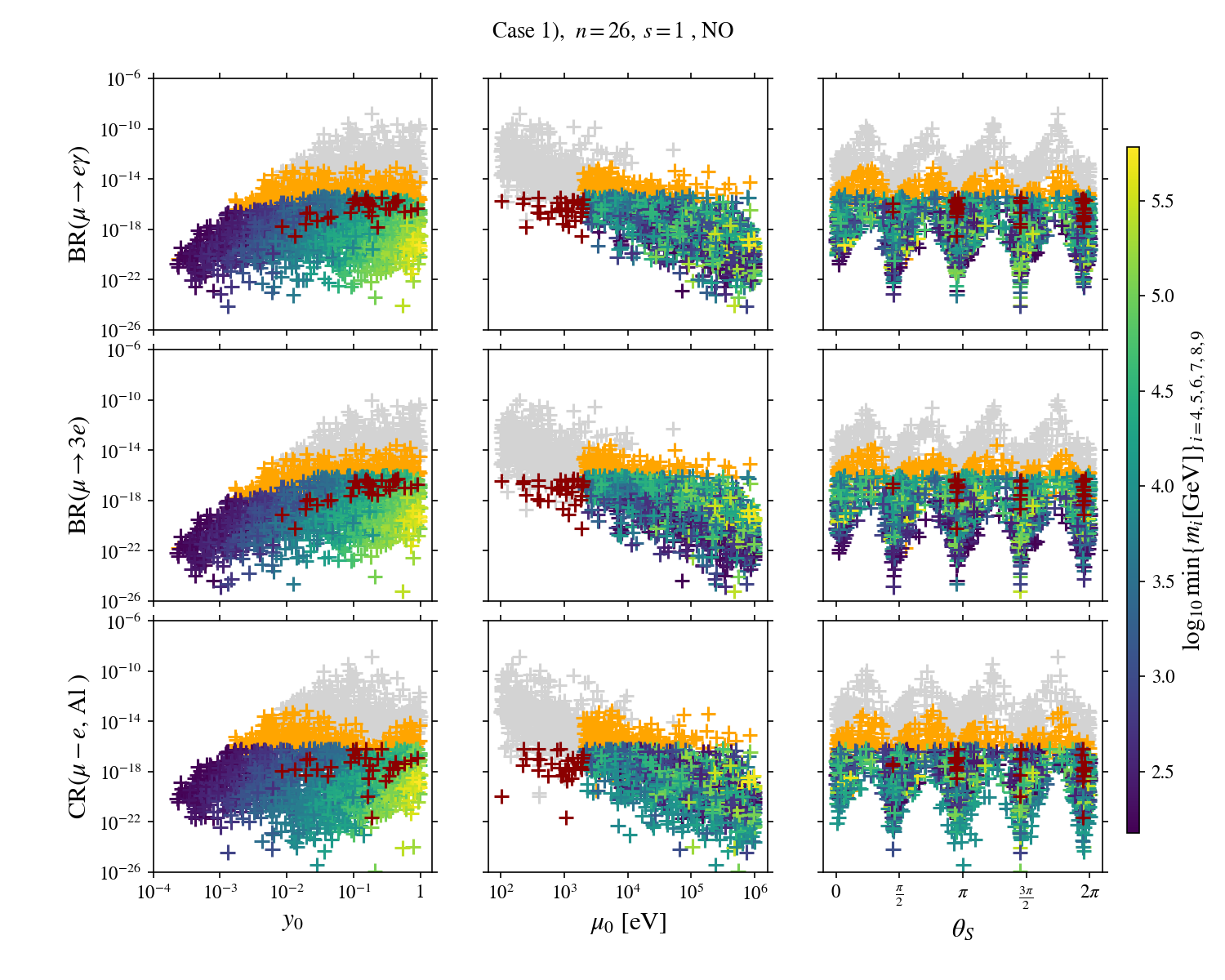}
    \caption{{\bf \mathversion{bold}Case 1). Results of numerical scan for $\mathrm{BR} (\mu\to e \, \gamma)$, $\mathrm{BR} (\mu\to 3 \,e)$ and $\mathrm{CR} (\mu-e, \mathrm{Al})$ varying $y_0$, $\mu_0$ and $\theta_S$\mathversion{normal}} 
    in the ranges in eqs.~(\ref{eq:y0range}), (\ref{eq:mu0range}) and (\ref{eq:thetaSrange}), respectively. The parameters $n$ and $s$ are the same as in fig.~\ref{fig:Case1mu0fixedNOIO}. The light neutrino mass ordering is fixed to NO and $m_0$
    to $m_0=0.03 \, \mathrm{eV}$. Points in grey, orange and red are excluded by different (current/future) experimental bounds, see text for details. Points in other colours correspond to a certain value of the lightest mass of the heavy sterile
    states, see colour bar, and pass all the imposed limits. 
    }
    \label{fig:Case1n26s1NOm0003}
\end{figure}

The colour-coding of the data points in the plots, e.g.~fig.~\ref{fig:Case1n26s1NOm0003}, is as follows: for grey points at least one of the future limits on $\mathrm{BR} (\mu\to e \gamma)$, 
$\mathrm{BR} (\mu\to 3\, e)$ or $\mathrm{CR} (\mu-e, \mathrm{Al})$
is violated as well as the current bound on at least one of the elements of the matrix $\eta$. In the case of orange points at least one of the future limits is violated, but all current constraints on the magnitude of the
matrix elements $\eta_{\alpha\beta}$ are fulfilled, while red points represent data points that are excluded by the current bounds on $\eta_{\alpha\beta}$, but not by the prospective limits on the three studied $\mu-e$ transitions. The different
colours that can be read off from the colour bar in the plot indicate the minimum of the masses $m_{4,...,9}$ of the heavy sterile states for each viable data point which satisfies both the expected bounds on the mentioned
cLFV processes and the current limits on $\eta_{\alpha\beta}$.

\paragraph{Case 1)} As representative we use
\begin{equation}
\label{eq:repCase1}
n=26 \;\; \mbox{and} \;\; s=1 \; ,
\end{equation}
noting that the particular choice of $s$, the CP symmetry, is not relevant. 
 The results for light neutrino masses with NO and $m_0=0.03 \, \mathrm{eV}$ are reported in fig.~\ref{fig:Case1n26s1NOm0003}. Those for IO light neutrino masses look very similar and, thus, are omitted.
We observe that the BR of $\mu\to e\gamma$ turns out to be smaller than about $6 \times 10^{-16}$, in case current bounds on the elements of the matrix $\eta$ and future limits on the three $\mu-e$ transitions are taken into account.
A lower limit does not seem to exist. The results for $\mathrm{BR} (\mu\to 3 \, e)$ show that also this BR can be (very) suppressed, while its maximum for viable data points (coloured according to the colour bar) is $10^{-16}$, meaning that
the prospective limit from Mu3E Phase 2 is effective. Analogous conclusions hold for $\mathrm{CR} (\mu-e, \mathrm{Al})$, i.e.~(exact) cancellation in the CR is possible and the maximum value for viable data points is determined by the expected bound coming from COMET. Furthermore, we find that the signal strength of the studied $\mu-e$ transitions is enhanced for $\theta_S$ satisfying $\cos 2 \, \theta_S \approx 0$. On the other hand, for $|\cos 2 \, \theta_S|$ close to one, the BRs and CR become (strongly) suppressed. Considering smaller values of the lightest neutrino mass $m_0$, e.g.~$m_0=10^{-5} \, \mathrm{eV}$, can lead to variations in the dependence of the BRs and CR on the angle $\theta_S$, compare fig.~\ref{fig:Case1n26s1IOm010em5} for light neutrino masses with IO in appendix~\ref{app:suppplots}. Additionally, it seems to be more unlikely to obtain strongly suppressed values of the two BRs.

\begin{figure}[t!]
    \centering
     \includegraphics[width=\textwidth]{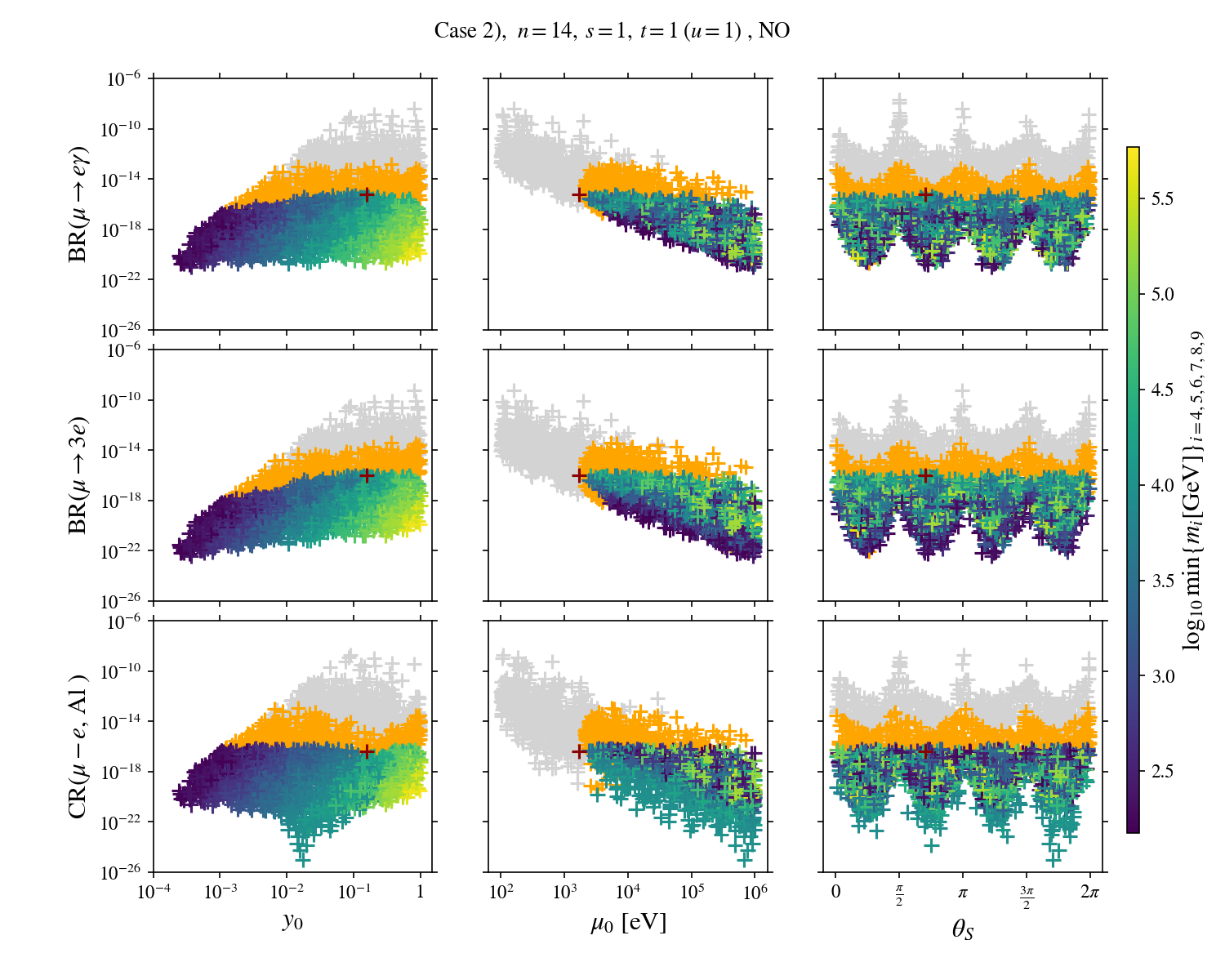}
    \caption{{\bf \mathversion{bold}Case 2). Results of numerical scan for $\mathrm{BR} (\mu\to e \, \gamma)$, $\mathrm{BR} (\mu\to 3 \,e)$ and $\mathrm{CR} (\mu-e, \mathrm{Al})$ varying $y_0$, $\mu_0$ and $\theta_S$\mathversion{normal}} 
    in the ranges in eqs.~(\ref{eq:y0range}), (\ref{eq:mu0range}) and (\ref{eq:thetaSrange}), respectively. The parameters $n$, $s$ and $t$ are set to $n=14$, $s=1$ and $t=1$ corresponding to $u=1$ (and $v=3$). 
    The light neutrino mass ordering is fixed to NO and $m_0$
    to $m_0=0.03 \, \mathrm{eV}$. The colour-coding of the data points is the same as in fig.~\ref{fig:Case1n26s1NOm0003}.
    }
    \label{fig:Case2n14s1t1NOm0003}
\end{figure}

\paragraph{Case 2)} We choose as representatives  
\begin{equation}
\label{eq:repCase2}
n=14 \;\; \mbox{and} \;\; (s,t)={ (1,1), (1,2), (0,1) }
\end{equation}
corresponding to $u=1, 0, -1$ (and $v=3, 6, 3$), respectively, since $s$ and $t$, that characterise the CP symmetry, are related to $u$ (and $v$) by $u=2 \, s-t$ (and $v=3 \, t$), cf.~\cite{Hagedorn:2014wha}.
The most striking difference between the various choices of $s$ and $t$ is the fact that for $t$ odd ($u$ odd) the BRs and CR reveal a dependence on the angle $\theta_S$, while this is not the case for
$t$ even ($u$ even). This is due to the fact that for $t$ even ($u$ even) the combination in eq.~(\ref{eq:comb}) is diagonal, whereas it is only block-diagonal for $t$ odd ($u$ odd), see also~\cite{Drewes:2022kap}.
\begin{figure}[t!]
    \centering
     \includegraphics[width=\textwidth]{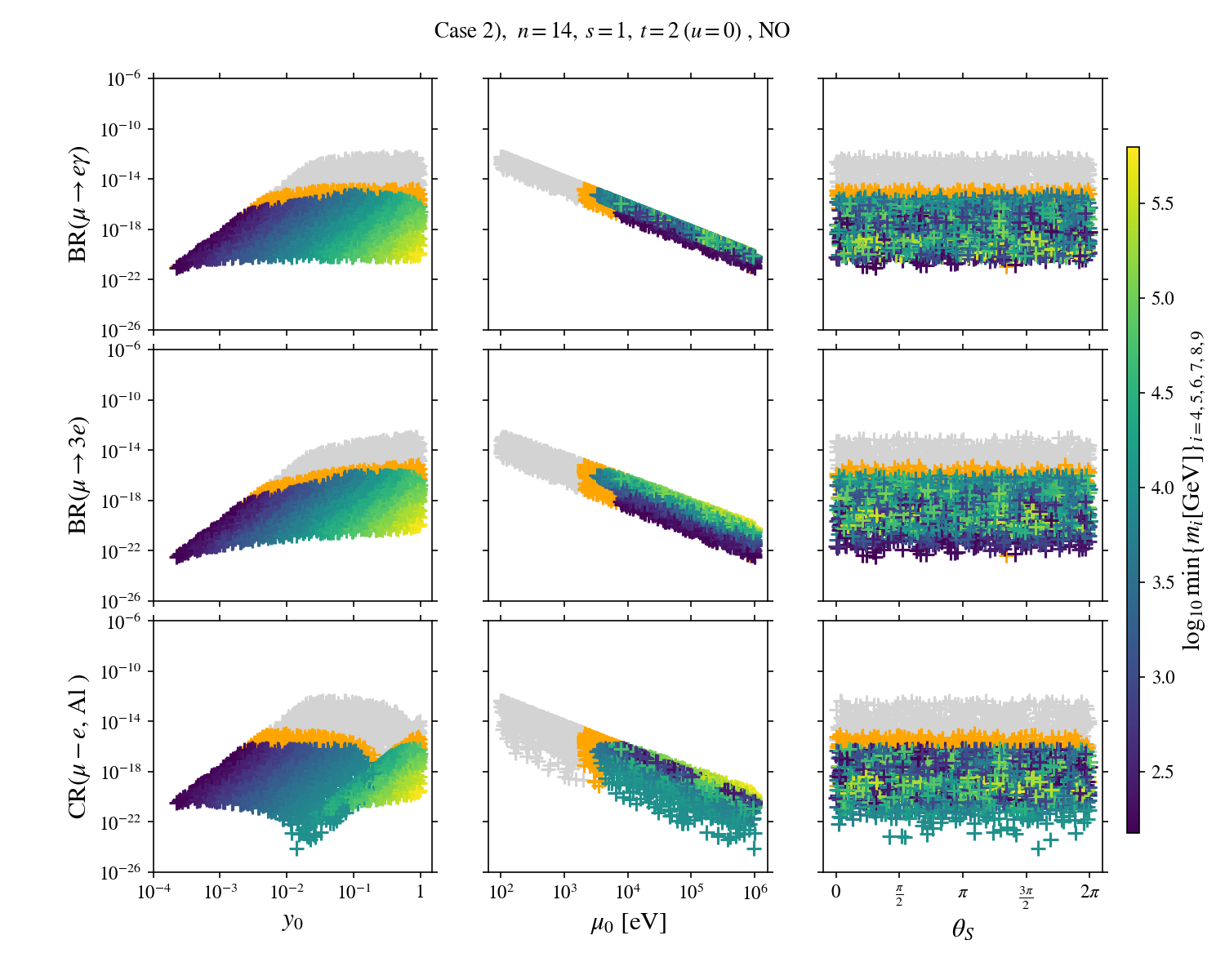}
    \caption{{\bf \mathversion{bold}Case 2). Results of numerical scan for $\mathrm{BR} (\mu\to e \, \gamma)$, $\mathrm{BR} (\mu\to 3 \,e)$ and $\mathrm{CR} (\mu-e, \mathrm{Al})$ varying $y_0$, $\mu_0$ and $\theta_S$\mathversion{normal}} 
    in the ranges in eqs.~(\ref{eq:y0range}), (\ref{eq:mu0range}) and (\ref{eq:thetaSrange}), respectively. The parameters $n$, $s$ and $t$ are set to $n=14$, $s=1$ and $t=2$ corresponding to $u=0$ (and $v=6$). 
    The light neutrino mass ordering is fixed to NO and $m_0$
    to $m_0=0.03 \, \mathrm{eV}$. The colour-coding of the data points is the same as in fig.~\ref{fig:Case1n26s1NOm0003}.
    }
    \label{fig:Case2n14s1t2NOm0003}
\end{figure}
Given this difference we display results for both $u=0$ (i.e.~for $s=1$ and $t=2$) and $u=1$ (i.e.~for $s=1$ and $t=1$) in figs.~\ref{fig:Case2n14s1t2NOm0003} and~\ref{fig:Case2n14s1t1NOm0003} 
for NO light neutrino masses and $m_0=0.03 \, \mathrm{eV}$. As one can clearly see, in the latter case the BRs and CR are enhanced for $\sin 2 \, \theta_S \approx 0$ and suppressed for the magnitude
of $\sin 2 \, \theta_S$ being close to one. This dependence on $\theta_S$ should be contrasted with the one observed for Case 1). Furthermore, we find that the largest possible values of the BRs and CR for
viable data points are very similar for Case 1) and Case 2), whereas for Case 2) $\mathrm{BR} (\mu \to e \, \gamma)$ and $\mathrm{BR} (\mu \to 3 \, e)$ appear to always have values larger than $10^{-21}$
and $10^{-23}$, respectively. For the choice $u=-1$ hold the same statements as for $u=1$. Furthermore, IO light neutrino masses lead to similar results. For a smaller value of $m_0$, $m_0=10^{-5} \, \mathrm{eV}$,
 the obtained results are similar to those presented, with the tendency to render suppressed values of the BRs and CR less likely.
 We note that in all plots both values of the angle $\theta_N$ that allow for a good fit of the measured lepton mixing angles are used and no difference between them is found, when scanning $y_0$, $\mu_0$ and $\theta_S$
  in the ranges in eqs.~(\ref{eq:y0range}), (\ref{eq:mu0range}) and (\ref{eq:thetaSrange}), respectively. Yet, in plots with a fixed value of $\mu_0$ slight differences are encountered, compare
 figs.~\ref{fig:Case2mu0fixedNOIO} and~\ref{fig:Case2mu0fixedsmallNOIO} in appendix~\ref{app:suppplots}. In such plots, also the fact that $\mathrm{CR} (\mu-e, \mathrm{Al})$ is only suppressed, 
 but there exists no exact cancellation for Case 2) can be seen, which is not visible well in the numerical scans.

\begin{figure}[t!]
    \centering
     \includegraphics[width=\textwidth]{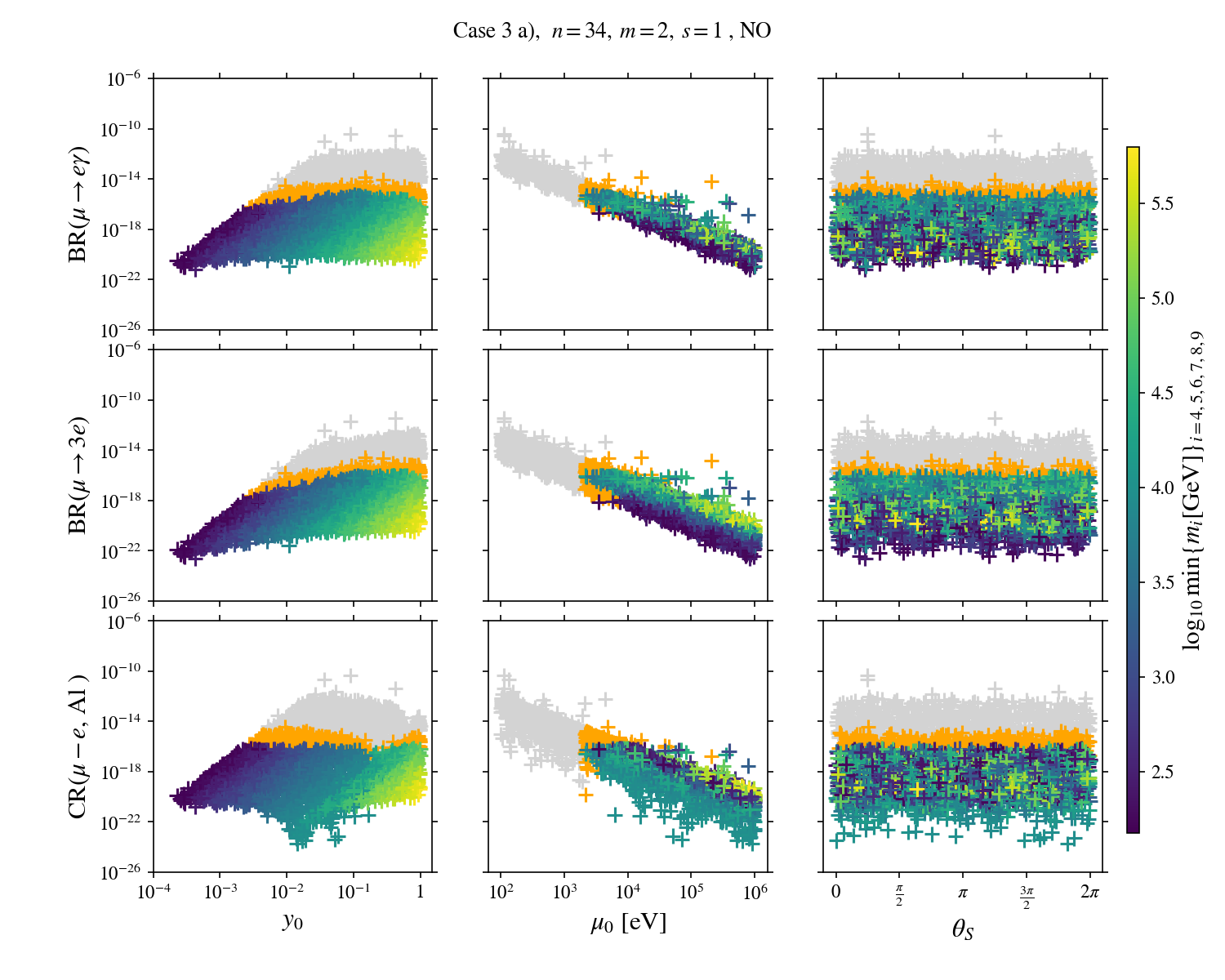}
    \caption{{\bf \mathversion{bold}Case 3 a). Results of numerical scan for $\mathrm{BR} (\mu\to e \, \gamma)$, $\mathrm{BR} (\mu\to 3 \,e)$ and $\mathrm{CR} (\mu-e, \mathrm{Al})$ varying $y_0$, $\mu_0$ and $\theta_S$\mathversion{normal}} 
    in the ranges in eqs.~(\ref{eq:y0range}), (\ref{eq:mu0range}) and (\ref{eq:thetaSrange}), respectively. The parameters $n$, $m$ and $s$ are set to $n=34$, $m=2$ and $s=1$. 
    The light neutrino mass ordering is fixed to NO and $m_0$
    to $m_0=0.03 \, \mathrm{eV}$. The colour-coding of the data points is the same as in fig.~\ref{fig:Case1n26s1NOm0003}. 
    }
    \label{fig:Case3an34s1m2NOm0003}
\end{figure}

\paragraph{Case 3 a)} Representatives of this case are
\begin{equation}
\label{eq:repCase3a}
n=34 \; , \;\; m=2 \;\; \mbox{and} \;\; s=1,2 \; . 
\end{equation}
Like for Case 2), we find that the combination of the parameters $m$ (determining the generator of the residual $Z_2$ symmetry among the neutral states) and $s$ (corresponding to the CP symmetry)
 fixes the dependence on the angle $\theta_S$, i.e.~for $m$ and $s$ being both even or odd the BRs and CR do not depend on $\theta_S$, whereas for $m$ even and $s$ odd (or vice versa, as possible for
 a different choice of the index $n$) the BRs and CR depend on $\cos 2 \, \theta_S$, meaning for $\cos 2 \, \theta_S \approx 0$ their values are enhanced and suppressed for $|\cos 2 \, \theta_S| \approx 1$. This
 is very similar to what is observed for Case 1). Since the plots for $m=2$ and $s=2$ and $m_0$ fixed as in eq.~(\ref{eq:m0benchmarks}) look similar to the analogous plots for Case 2) ($n=14$, $s=1$, $t=2$), see
  fig.~\ref{fig:Case2n14s1t2NOm0003}, we refrain from displaying them. Instead, we show in fig.~\ref{fig:Case3an34s1m2NOm0003} results for $m=2$ and $s=1$ and 
  NO light neutrino masses with $m_0=0.03 \, \mathrm{eV}$. Although a dependence on the 
 angle $\theta_S$ is expected, it is hardly visible, compared to Case 1) in fig.~\ref{fig:Case1n26s1NOm0003} and Case 2) in fig.~\ref{fig:Case2n14s1t1NOm0003}. 
 When considering light neutrino masses with IO similar results are found.
Only for smaller values of $m_0$, $m_0=10^{-5} \, \mathrm{eV}$, and NO light neutrino masses the dependence of the BRs and CR on $\theta_S$ becomes more pronounced, 
see fig.~\ref{fig:Case3an34s1m2NOm010em5} in appendix~\ref{app:suppplots}.
In addition, we note the peculiar feature that two branches in the scan results are visible (for small values of $y_0$) which correspond to the two different choices of the angle $\theta_N$ that lead to an 
agreement with the measured values of the lepton mixing angles at the $3 \, \sigma$ level or better. This is best seen in fig.~\ref{fig:Case3an34s2m2NOm010em5}, also found in the appendix.
 Regarding the size of the BRs and CR, like for Case 2) the BRs typically observe as lower limits $\mathrm{BR} (\mu\to e \, \gamma) \gtrsim 10^{-21}$ and $\mathrm{BR} (\mu\to 3 \, e) \gtrsim 10^{-23}$, while the CR can, in
 principle, be very suppressed. Furthermore, we find for the viable data points very similar maximal values that the BRs and CR attain as in Case 1) and Case 2).

\paragraph{Case 3 b.1)} For this case we have analysed different combinations with
\begin{equation}
\label{eq:repCase3b1}
n=20 \; , \;\; m=9,10,11 \;\; \mbox{and} \;\; 0 \leq s \leq 19 \; ,
\end{equation}
always ensuring that they can describe the measured lepton mixing angles at the $3 \, \sigma$ level or better for at least one value of the angle $\theta_N$.
As expected, we find for combinations $m$ and $s$ both even or odd no dependence of the BRs and CR on the angle $\theta_S$, while for $m$ being even and $s$ odd or vice versa
such a dependence exists and is (mainly) determined by $|\cos 2 \, \theta_S|$. Unlike for Case 3 a), this dependence is visible well, similar to the findings for Case 1), see fig.~\ref{fig:Case1n26s1NOm0003}.
We have, furthermore, checked that results for light neutrino masses with IO are in general similar to those for NO light neutrino masses.
For smaller values of $m_0$, we can, depending on the choice of the parameters $m$ and $s$, find a variation of the dependence of the BRs and CR on $\theta_S$. This is similar to what
is reported for Case 1), compare figs.~\ref{fig:Case1n26s1NOm0003} and~\ref{fig:Case1n26s1IOm010em5}.
 For most of the studied combinations of $m$ and $s$ the BRs of $\mu\to e \, \gamma$ and $\mu\to 3 \, e$ appear to be larger than $10^{-22}$ and $10^{-23}$, respectively, while for e.g.~$m=11$ and $s=0$ their size can be 
 more suppressed. For $\mathrm{CR} (\mu-e, \mathrm{Al})$ minimum values usually smaller than $10^{-24}$ can be observed. Regarding the maximal values of the BRs and CR that are found for viable data
 points, these are similar to those reported for the other cases.

\subsection{Comments on cLFV tau lepton decays}
\label{taucLFVs}

%
\begin{figure}[t!]
    \centering
     \includegraphics[width=\textwidth]{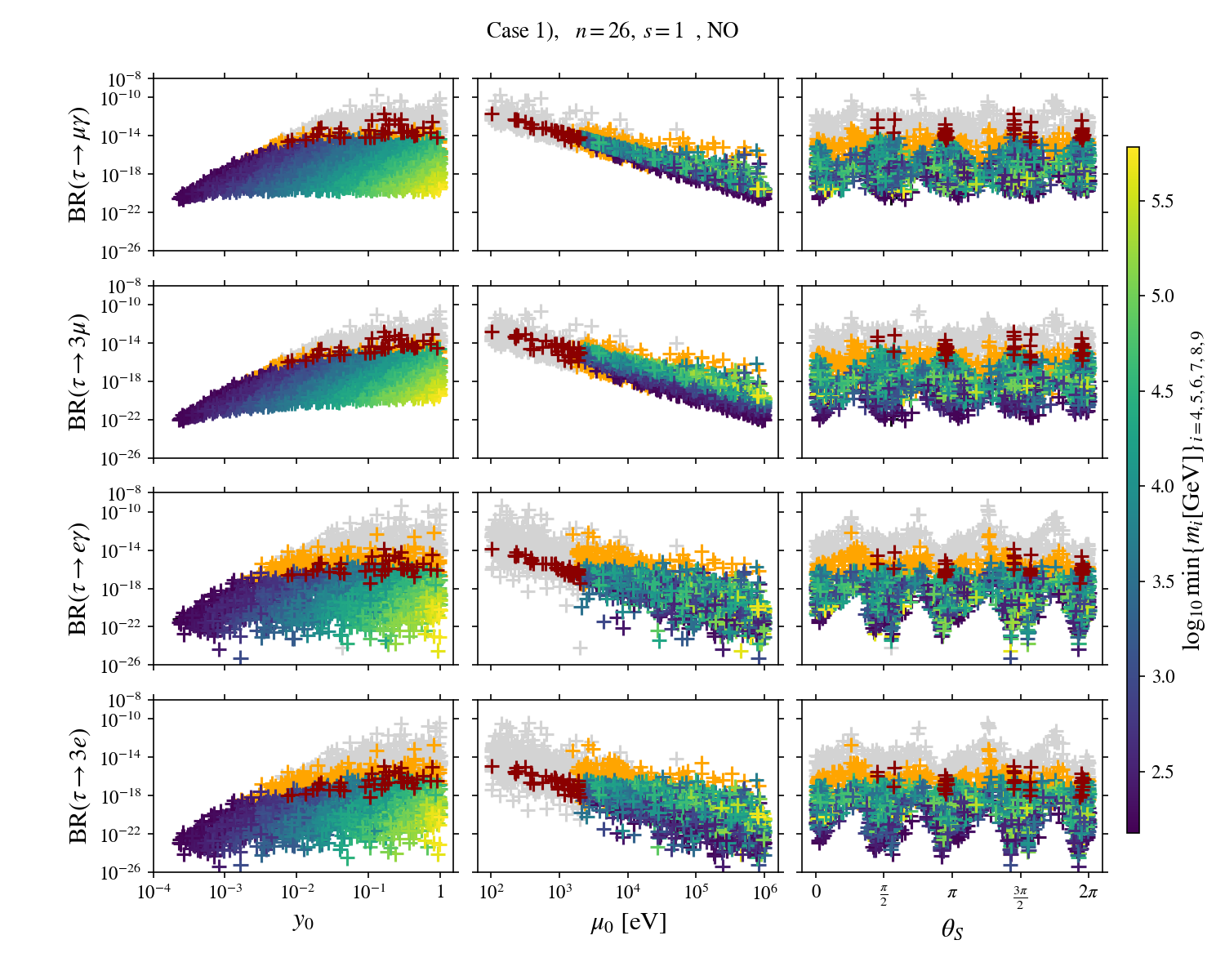}
    \caption{{\bf \mathversion{bold}Case 1). Results of numerical scan for $\mathrm{BR} (\tau\to \mu \, \gamma)$, $\mathrm{BR} (\tau\to 3 \,\mu)$, $\mathrm{BR} (\tau\to e \, \gamma)$ and $\mathrm{BR} (\tau\to 3 \,e)$ varying 
    $y_0$, $\mu_0$ and $\theta_S$\mathversion{normal}} 
    in the ranges in eqs.~(\ref{eq:y0range}), (\ref{eq:mu0range}) and (\ref{eq:thetaSrange}), respectively. The parameters $n$ and $s$ are chosen as $n=26$ and $s=1$. 
    The light neutrino mass ordering is fixed to NO and $m_0$
    to $m_0=0.03 \, \mathrm{eV}$. The colour-coding of the data points is the same as in fig.~\ref{fig:Case1n26s1NOm0003}. 
    }
    \label{fig:Case1n26s1NO_taudecays}
\end{figure}
In this section, we briefly comment on the BRs of the four cLFV tau lepton decays $\mathrm{BR} (\tau\to \mu \, \gamma)$, $\mathrm{BR} (\tau\to 3 \,\mu)$, $\mathrm{BR} (\tau\to e \, \gamma)$ and $\mathrm{BR} (\tau\to 3 \,e)$.
For viable data points (taking into account the current limits on the matrix elements $\eta_{\alpha\beta}$ as well as the future bounds on the three studied $\mu-e$ transitions), the obtained BRs are in general no larger than $10^{-14}$
and, thus, too small to be observed at Belle~II. In most of the analysed situations they are bounded from below by $10^{-23}$.
For this reason, we just display one set of plots for Case 1), $n=26$ and $s=1$, assuming that light neutrino masses follow NO and $m_0=0.03 \, \mathrm{eV}$, 
see fig.~\ref{fig:Case1n26s1NO_taudecays}. The plots for IO light neutrino masses look similar. Like for the $\mu-e$ transitions, compare fig.~\ref{fig:Case1n26s1NOm0003}, 
also the BRs of the cLFV tau lepton decays possess a dependence on the angle $\theta_S$. An analogous statement holds for the representatives of the other cases, Case 2) through Case 3 b.1).
Furthermore, we note that for the displayed case, the BRs of $\tau\to e \, \gamma$ and $\tau\to 3 \, e$ can be more suppressed than those
for $\tau\to \mu \, \gamma$ and $\tau\to 3 \, \mu$. Apart from this type of feature, in fig.~\ref{fig:Case3an34s2m2NOm010em5_taudecays} in appendix~\ref{app:suppplots} 
one can observe for Case 3 a), $n=34$, $m=2$, $s=2$, light neutrino masses with NO and $m_0=10^{-5} \, \mathrm{eV}$ that two branches
become (slightly) visible (for small $y_0$), when studying the BRs of $\tau\to e \gamma$ and $\tau\to 3 \, e$. Like for the analysed $\mu-e$ transitions, compare figs.~\ref{fig:Case3an34s1m2NOm010em5} and~\ref{fig:Case3an34s2m2NOm010em5} 
in the appendix, these branches are associated with the two different values of the angle $\theta_N$ which allow to accommodate the lepton mixing angles at the $3 \, \sigma$ level or better. However, such
branches cannot be seen in the results for $\mathrm{BR} (\tau\to \mu \, \gamma)$ and $\mathrm{BR} (\tau\to 3 \,\mu)$.

\section{Comparison between the different options}
\label{comparison}

We compare option 3 and its results, studied in this work, with those for option 2, analysed in~\cite{DiMeglio:2024gve} as well as the ones for option 1, see~\cite{Hagedorn:2021ldq}.
 
Firstly, we remind that the assignment of the different fields under the flavour symmetry can differ for the three options. While LH lepton doublets and RH charged leptons are assigned to the same irreps for all three options,  
this is not the case for the sterile states $N_i$ and $S_j$, since they either transform in the same way (for option 1 both as ${\bf 3}$, while for option 2 both as ${\bf 3^\prime}$) or as different irreps, 
$N_i \sim {\bf 3}$ and $S_j \sim {\bf 3^\prime}$, for option 3. This is necessary in order to achieve the desired flavour structures of the mass matrices $m_D$, $M_{NS}$ and $\mu_S$.
Furthermore, all fields are assigned to the same charges under $Z_3^{\mathrm{(aux)}}$ for the three options.
 
When comparing option 2 and 3, we note that the form of the matrix $M_{NS}$ in eq.~(\ref{eq:formMNS}) for option 3 is very similar to the one of $m_D$ for option 2
 and that the distinction between having the combination in eq.~(\ref{eq:comb}) diagonal or only block-diagonal is similar to the discussion for option 2; see also~\cite{Drewes:2022kap} for a setup with type-I
seesaw. Indeed, if this combination is diagonal, the result for lepton mixing, compare eq.~(\ref{eq:UnutcasediagLO}), is analogous to that obtained for the other two options. 
 In case it is only block-diagonal, the angle $\bar{\theta}_N$, defined below eq.~(\ref{eq:mfMf_Case3b1}), corresponds to the effective angle $\bar{\theta}_L$ for option 2.
 Also the effects of non-unitarity (in lepton mixing), see eqs.~(\ref{eq:nonuni_diag}) and (\ref{eq:nonuni_nondiag}), for option 3 are qualitatively similar to the ones observed for option 2, whereas
 for option 1 the matrix $\eta$ has a flavour-diagonal and -universal form, leading to the same suppression of all elements of the matrix $U_0$, compare eq.~(\ref{eq:UPMNSetaU0}). 

A striking difference between option 3 and the other two options lies in the mass spectrum of the heavy sterile states. While these are all (nearly) degenerate in mass for option 1 and 2,
they form three pseudo-Dirac pairs with in general different masses for option 3, see eqs.~(\ref{eq:heavymasses}-\ref{eq:m49MfnondiagCase1}) and fig.~\ref{fig:Mfm0NOIO}. For light neutrino masses
following IO, nevertheless, two of these pairs become close in mass, cf.~eq.~(\ref{eq:miratiosIOm0large}). In addition, due to the choice of the parameter space for option 3, see section~\ref{prerequisites},
the masses of the heavy sterile states can vary in a larger range in the numerical scans; in particular, they can be larger (depending on the lightest neutrino mass $m_0$), while for option 1 and 2 the mass parameter 
corresponding to the mass scale of the heavy sterile states is always chosen to be no larger than $10 \, \mathrm{TeV}$.

Another distinctive feature of option 2 and 3 is the prediction of signal strengths of $\mu\to e \, \gamma$, $\mu\to 3 \, e$ and $\mu-e$ conversion in aluminium that are in the reach of current/near-future experiments, 
whereas it has been shown that for option 1 all cLFV signals are very suppressed like in the SM with light neutrino masses. Indeed, the expected rates of the studied cLFV processes,
$\mathrm{BR} (\mu\to e \, \gamma)$, $\mathrm{BR} (\mu \to 3 \, e)$, $\mathrm{CR} (\mu-e, \mathrm{Al})$, $\mathrm{BR} (\tau\to \mu \, \gamma)$, $\mathrm{BR} (\tau\to 3 \,\mu)$, 
$\mathrm{BR} (\tau\to e \, \gamma)$ and $\mathrm{BR} (\tau\to 3 \,e)$, are very similar for option 2 and 3, taking into account the fact that for option 3 the masses of the heavy sterile states
can be larger than for option 2 in the numerical scans. Also features, observed for the different cases, Case 1) through Case 3 b.1), are found for both options, e.g.~the fact that 
for Case 1) the results are independent of the parameter $s$ that characterises the CP symmetry, compare section~\ref{muetransitions}. In particular, we see that the dependence of the different BRs and 
the CR on the angle $\theta_S$ encountered when studying option 3, see e.g.~fig~\ref{fig:Case1n26s1NOm0003}, is very similar to the dependence on the angle $\theta_R$ for option 2, for all considered cases. 
 Furthermore, the minimum value of $\mu_0$ necessary in order to pass the current bounds on the matrix elements $\eta_{\alpha\beta}$ and the prospective limits on the three analysed $\mu-e$ transitions is typically 
 bounded from below by $2 \, \mathrm{keV}$ for both options 2 and 3.
  
However, there are also differences among the results for option 2 and 3. Probably, the most noteworthy one is related to the possible cancellation in $\mathrm{CR} (\mu-e, \mathrm{Al})$.
 For option 2, such a cancellation always occurs and the value of the mass of the heavy sterile states at which it occurs is universal, i.e.~it does not depend on the case, Case 1) through Case 3 b.1), 
 but only on the choice of the nucleus. For option 3, instead we have found that such an exact cancellation does not always exist, e.g.~for Case 2) (compare figs.~\ref{fig:Case2mu0fixedNOIO} 
 and \ref{fig:Case2mu0fixedsmallNOIO} in appendix~\ref{app:suppplots}), and if it exists, it can depend on the 
 chosen case, the group theory parameters as well as $\mu_0$ and the angles $\theta_N$ and $\theta_S$. So, a measurement of $\mathrm{CR} (\mu-e, \mathrm{Al})$
 in combination with a determination of the mass of (some of) the heavy sterile states could constrain the viable options and/or cases. Additionally, we remark that, in principle, it is possible
 to also observe a cancellation in one of the BRs for a certain combination of parameters for option 3, compare comments at the end of section~\ref{analyticscLFV}, while this is not the case for option 2. 
 Eventually, for option 3 we can find in some instances two visible branches due to the two different viable values of the angle $\theta_N$, see the numerical scan for Case 3 a), NO light neutrino masses and $m_0=10^{-5} \, \mathrm{eV}$ in 
 figs.~\ref{fig:Case3an34s1m2NOm010em5}-\ref{fig:Case3an34s2m2NOm010em5_taudecays} in appendix~\ref{app:suppplots}. This has not been observed in the analysis of option 2.

\section{Summary}
\label{summ}

We have considered an extension of the SM with 3+3 gauge singlets $N_i$ and $S_j$. Light neutrino masses are generated with the ISS mechanism.
Lepton mixing is determined by a flavour and CP symmetry and their residual groups among charged leptons and the neutral states and, consequently, predicted in terms of one real parameter.
 In the chosen basis, the charged lepton mass matrix is diagonal and contains three free parameters corresponding to the masses of the electron, muon and tau lepton. Among the neutral states,
  only the Dirac mass matrix $M_{N S}$ of the gauge singlets $N_i$ and $S_j$ has a non-trivial flavour structure that is fixed by the residual group $G_\nu = Z_2 \times CP$. 

 There are in total seven real parameters for the neutral states: three mass parameters $M_f$, $f=1,2,3$, $\mu_0$ the scale of lepton number violation, one Yukawa coupling $y_0$ and the angles $\theta_N$ and $\theta_S$.
 The light neutrino masses are adjusted with the help of $M_f$, while $\theta_N$ is used in order to achieve lepton mixing angles that agree at the $3 \, \sigma$ level or better with experimental data.
  The three remaining parameters, $\mu_0$, $y_0$ and $\theta_S$, are varied in well-motivated ranges in the numerical scans.
The heavy sterile states form three pairs of pseudo-Dirac particles with in general distinct masses ranging between $150 \, \mathrm{GeV}$ and $700 \, \mathrm{TeV}$, with the upper limit depending on the value of the lightest 
neutrino mass $m_0$. 

In the phenomenological study, we have focussed on the rates of different cLFV processes, in particular $\mu\to e \, \gamma$, $\mu\to 3 \, e$ and $\mu-e$ conversion in nuclei.
 While current bounds on these and the existing limits on the matrix elements $\eta_{\alpha\beta}$ do not considerably constrain the parameter space, prospective bounds on $\mathrm{BR} (\mu \to 3 \, e)$ 
 and $\mathrm{CR} (\mu-e, \mathrm{Al})$ impose relevant restrictions. Depending on the case (type of lepton mixing pattern) and the chosen parameters, 
 an exact cancellation in  $\mathrm{CR} (\mu-e, \mathrm{Al})$ is possible. In any case, in all analysed examples this CR can reach values as small as $10^{-24}$. 
 
Finally, we have compared the results for this variant of the ISS scenario (called option 3) with two other ones (option 1 and 2), studied already in the literature~\cite{Hagedorn:2021ldq,DiMeglio:2024gve}, and commented on similarities and 
important differences. On the one hand, option 1 and 2 predict heavy sterile states that are all (nearly) degenerate in mass, which is not the case for option 3. On the other hand, option 1 leads to very suppressed rates of the cLFV processes,
whereas option 2 and 3 can induce sizeable signals. The cancellation in $\mathrm{CR} (\mu-e, \mathrm{Al})$ is exact and universal for option 2 (only depending on the chosen nucleus), but not for option 3. It might, thus, be feasible to at least partly distinguish between
these different variants experimentally.

\section*{Acknowledgements}

We thank Miguel G.~Folgado for help with the local computer cluster.  This work has been/is supported by the Spanish MINECO through the Ram\'o{}n y Cajal programme  RYC2018-024529-I, the FPI fellowship PRE2021-098730, by the national grant PID2023-148162NB-C21, by the Generalitat Valenciana through PROMETEO/2021/083 as well as by the European Union's Horizon 2020 research and innovation programme under the Marie Skłodowska-Curie Staff Exchange grant agreement No.~101086085 (ASYMMETRY).

\appendix

\mathversion{bold}
\section{Supplementary plots}
\mathversion{normal}
\label{app:suppplots}

\begin{figure}[t!]
    \centering
     \includegraphics[width=\textwidth]{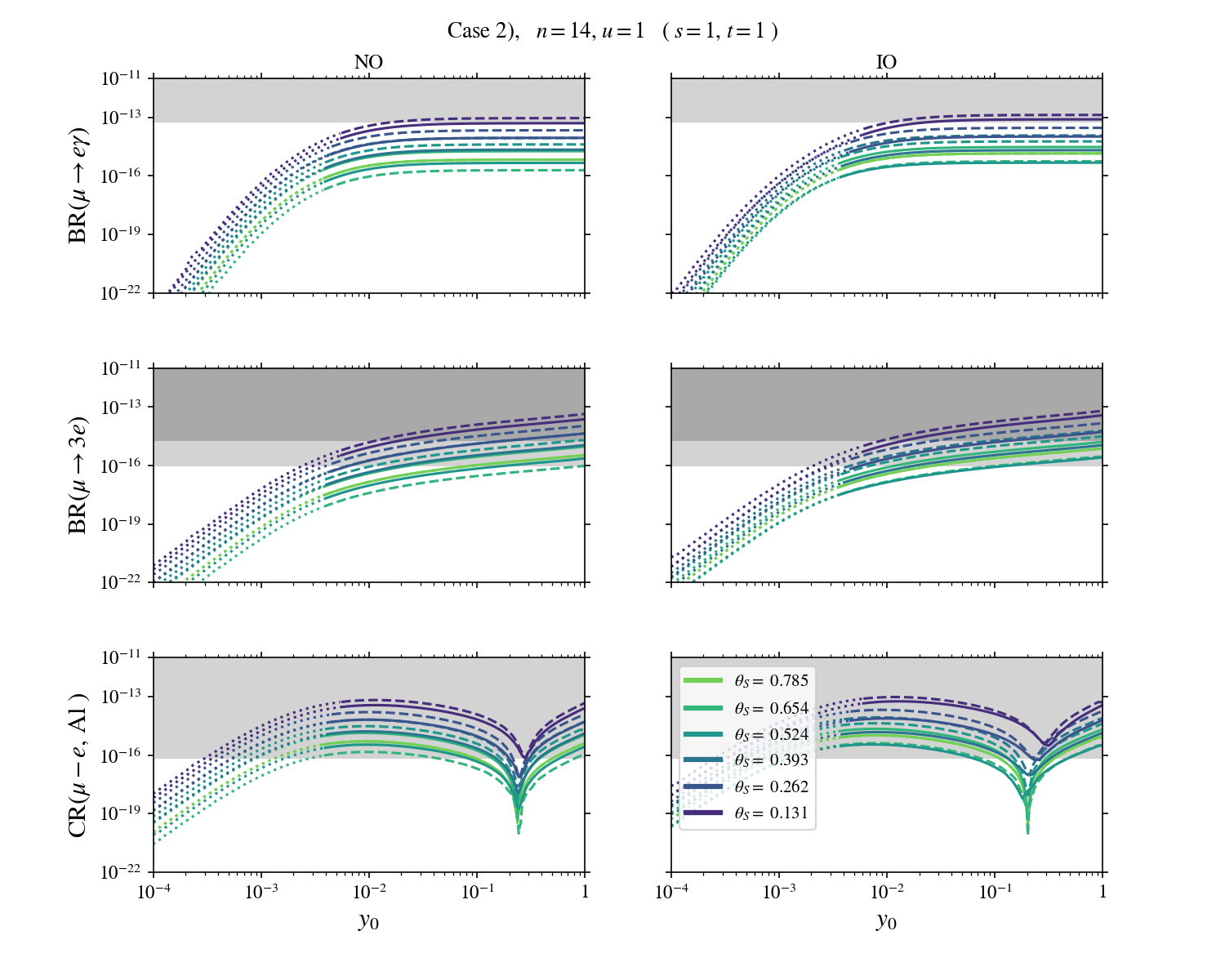}
    \caption{{\bf \mathversion{bold}Case 2). Results for $\mathrm{BR} (\mu\to e \, \gamma)$, $\mathrm{BR} (\mu\to 3 \,e)$ and $\mathrm{CR} (\mu-e, \mathrm{Al})$\mathversion{normal}} in the upper, middle and lower row, respectively. 
  These results are for $n=14$, $s=1$ and $t=1$ corresponding to $u=1$ and $v=3$, 
  and six different values of the angle $\theta_S$, shown in different colours. Results for the two different values of the angle $\theta_N$ that lead to a good agreement with the 
  experimental data on lepton mixing are shown with straight and dashed lines, respectively.
Otherwise, the conventions are the same as in fig.~\ref{fig:Case1mu0fixedNOIO}.
    }
    \label{fig:Case2mu0fixedNOIO}
\end{figure}

\begin{figure}[t!]
    \centering
     \includegraphics[width=\textwidth]{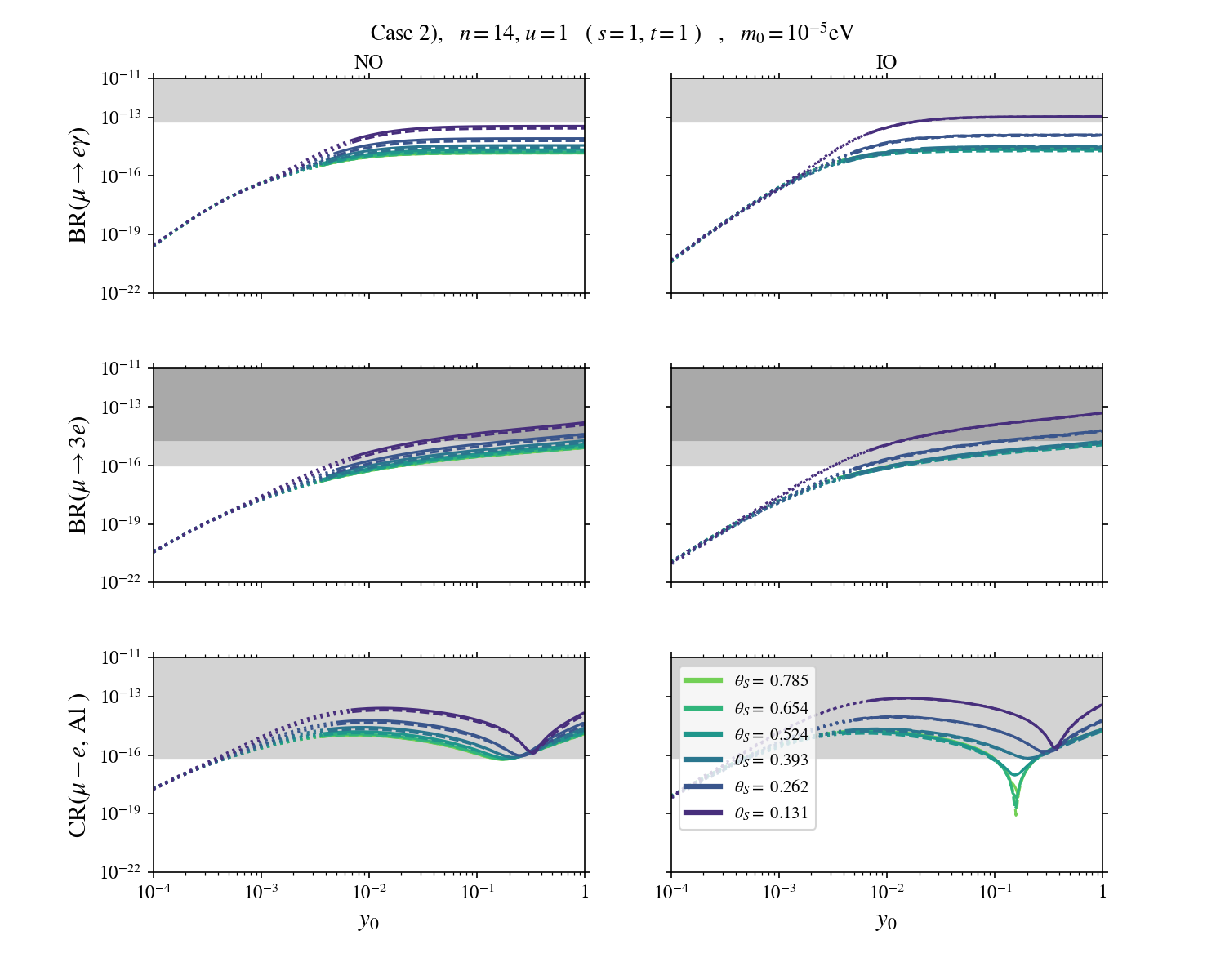}
    \caption{{\bf \mathversion{bold}Case 2). Results for $\mathrm{BR} (\mu\to e \, \gamma)$, $\mathrm{BR} (\mu\to 3 \,e)$ and $\mathrm{CR} (\mu-e, \mathrm{Al})$\mathversion{normal}} in the upper, middle and lower row, respectively. 
The lightest neutrino mass $m_0$ is chosen as $m_0=10^{-5} \, \mathrm{eV}$. The remaining conventions are the same as in fig.~\ref{fig:Case2mu0fixedNOIO}.
      }
    \label{fig:Case2mu0fixedsmallNOIO}
\end{figure}

In this appendix, we display supplementary plots. Figs.~\ref{fig:Case2mu0fixedNOIO} and~\ref{fig:Case2mu0fixedsmallNOIO} show results for the BRs of $\mu\to e \, \gamma$ and $\mu\to 3 \, e$
as well as the CR of $\mu-e$ conversion in aluminium for Case 2) and $n=14$, $s=1$ and $t=1$, corresponding to $u=1$ and $v=3$. We use the maximal value of the lightest neutrino mass $m_0$ that is compatible with
cosmological measurements in fig.~\ref{fig:Case2mu0fixedNOIO}, while fixing $m_0$ to $m_0=10^{-5} \, \mathrm{eV}$ in fig.~\ref{fig:Case2mu0fixedsmallNOIO}. 
 The plots for $n=14$, $s=0$ and $t=1$, corresponding to $u=-1$ and $v=3$, look similar. 
With these figures we further illustrate the statements made in section~\ref{analyticscLFV}, where we discuss the BRs and CR analytically.

\begin{figure}[t!]
    \centering
     \includegraphics[width=\textwidth]{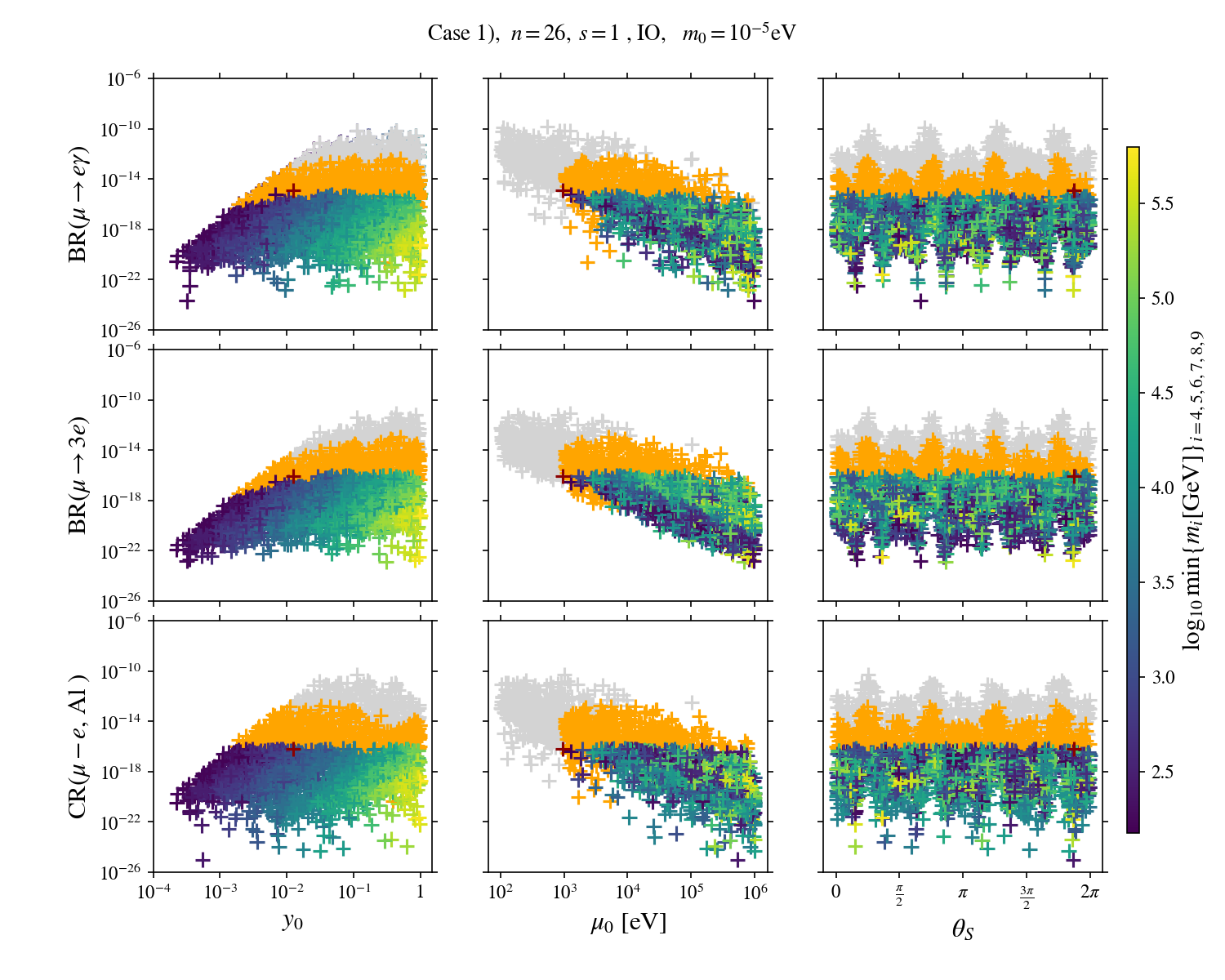}
    \caption{{\bf \mathversion{bold}Case 1). Results of numerical scan for $\mathrm{BR} (\mu\to e \, \gamma)$, $\mathrm{BR} (\mu\to 3 \,e)$ and $\mathrm{CR} (\mu-e, \mathrm{Al})$ varying $y_0$, $\mu_0$ and $\theta_S$\mathversion{normal}} 
    in the ranges in eqs.~(\ref{eq:y0range}), (\ref{eq:mu0range}) and (\ref{eq:thetaSrange}), respectively. The parameters $n$ and $s$ are the same as in fig.~\ref{fig:Case1mu0fixedNOIO}. The light neutrino mass ordering is fixed to IO and $m_0$
    to $m_0=10^{-5} \, \mathrm{eV}$. The rest of the conventions is the same as in fig.~\ref{fig:Case1n26s1NOm0003}. 
    }
    \label{fig:Case1n26s1IOm010em5}
\end{figure}

Fig.~\ref{fig:Case1n26s1IOm010em5} displays numerical results for the BRs and the CR for Case 1), $n=26$ and $s=1$, assuming that light neutrino masses follow IO and the lightest neutrino mass is set to $m_0=10^{-5} \, \mathrm{eV}$. 
This figure complements the results that are shown in the main text in fig.~\ref{fig:Case1n26s1NOm0003}.

\begin{figure}[t!]
    \centering
     \includegraphics[width=\textwidth]{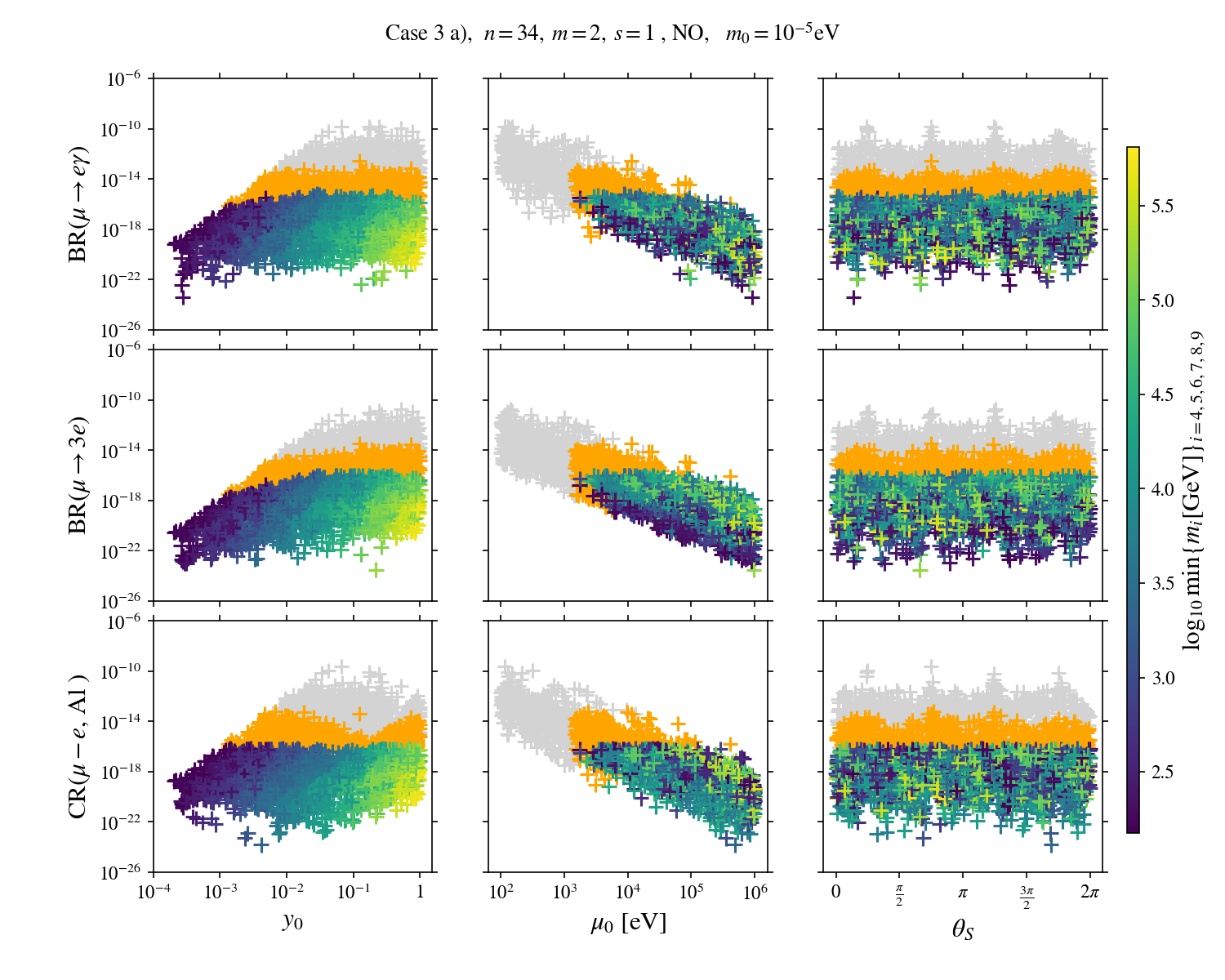}
    \caption{{\bf \mathversion{bold}Case 3 a). Results of numerical scan for $\mathrm{BR} (\mu\to e \, \gamma)$, $\mathrm{BR} (\mu\to 3 \,e)$ and $\mathrm{CR} (\mu-e, \mathrm{Al})$ varying $y_0$, $\mu_0$ and $\theta_S$\mathversion{normal}} 
    in the ranges in eqs.~(\ref{eq:y0range}), (\ref{eq:mu0range}) and (\ref{eq:thetaSrange}), respectively. The parameters $n$, $m$ and $s$ are set to $n=34$, $m=2$ and $s=1$. 
    The light neutrino mass ordering is fixed to NO and $m_0$
    to $m_0=10^{-5} \, \mathrm{eV}$. The colour-coding of the data points is the same as in fig.~\ref{fig:Case1n26s1NOm0003}. 
    }
    \label{fig:Case3an34s1m2NOm010em5}
\end{figure}

\begin{figure}[t!]
    \centering
     \includegraphics[width=\textwidth]{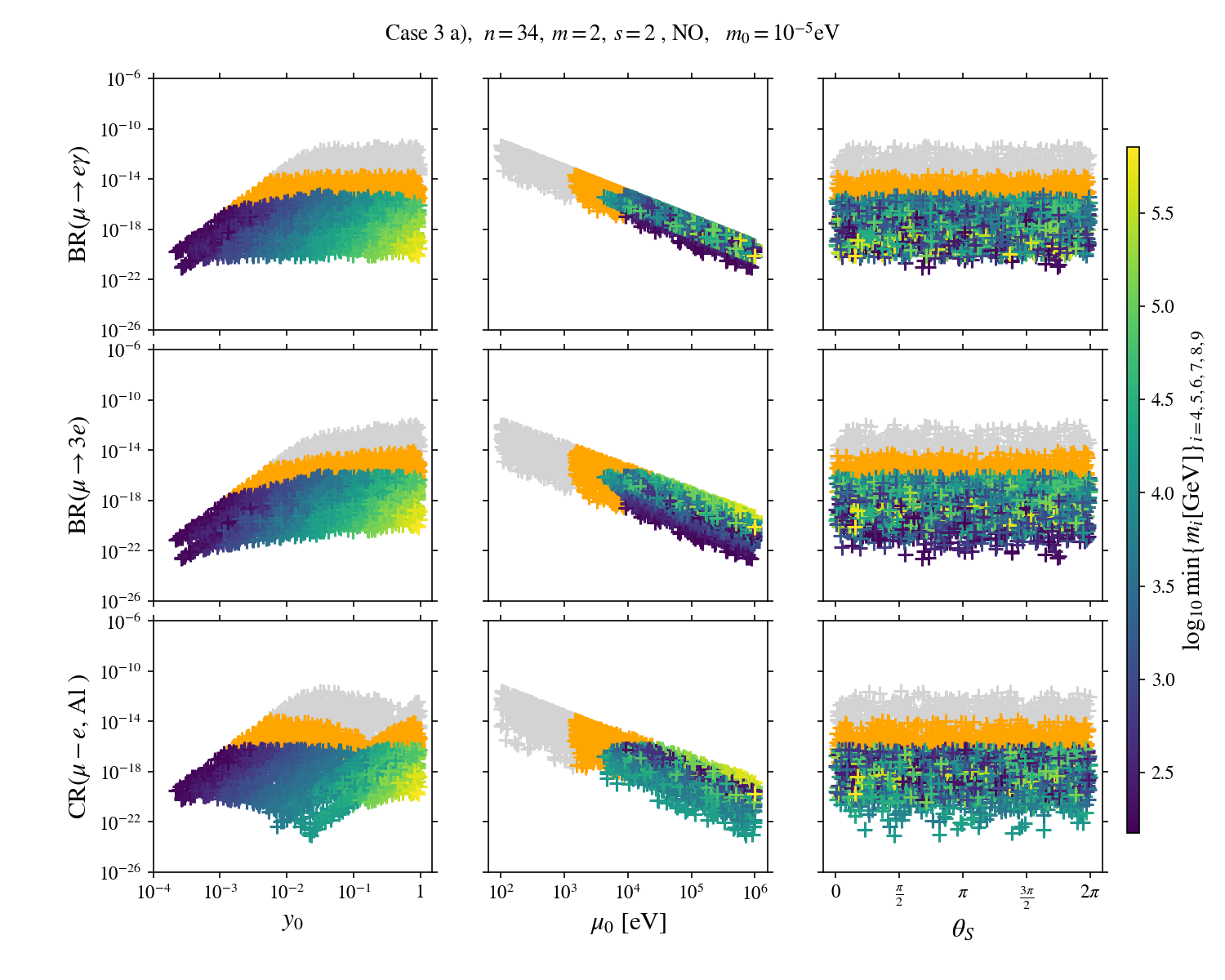}
    \caption{{\bf \mathversion{bold}Case 3 a). Results of numerical scan for $\mathrm{BR} (\mu\to e \, \gamma)$, $\mathrm{BR} (\mu\to 3 \,e)$ and $\mathrm{CR} (\mu-e, \mathrm{Al})$ varying $y_0$, $\mu_0$ and $\theta_S$\mathversion{normal}} 
    in the ranges in eqs.~(\ref{eq:y0range}), (\ref{eq:mu0range}) and (\ref{eq:thetaSrange}), respectively. The parameters $n$, $m$ and $s$ are set to $n=34$, $m=2$ and $s=2$. 
    The light neutrino mass ordering is fixed to NO and $m_0$
    to $m_0=10^{-5} \, \mathrm{eV}$. The colour-coding of the data points is the same as in fig.~\ref{fig:Case1n26s1NOm0003}. 
    }
    \label{fig:Case3an34s2m2NOm010em5}
\end{figure}

Figs.~\ref{fig:Case3an34s1m2NOm010em5} and~\ref{fig:Case3an34s2m2NOm010em5} show results of numerical scans for the BRs and the CR for Case 3 a), $n=34$, $m=2$ and $s=1$ and $s=2$, respectively, assuming that light neutrino masses follow NO and the lightest neutrino mass is set to $m_0=10^{-5} \, \mathrm{eV}$. These figures complete the discussion in the main text, in particular the results displayed in fig.~\ref{fig:Case3an34s1m2NOm0003}.

\begin{figure}[t!]
    \centering
     \includegraphics[width=\textwidth]{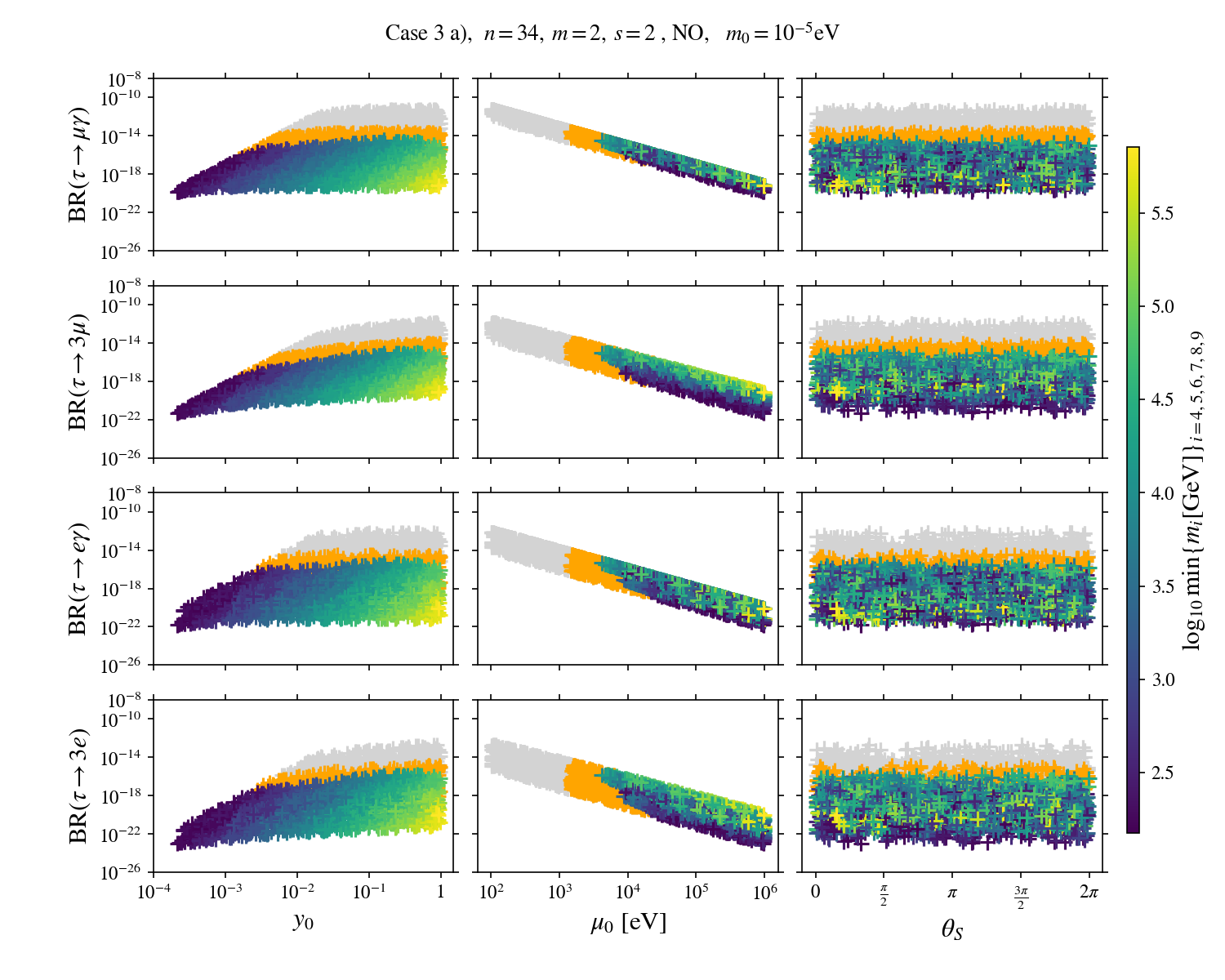}
    \caption{{\bf \mathversion{bold}Case 3 a). Results of numerical scan for $\mathrm{BR} (\tau\to \mu \, \gamma)$, $\mathrm{BR} (\tau\to 3 \,\mu)$, $\mathrm{BR} (\tau\to e \, \gamma)$ and $\mathrm{BR} (\tau\to 3 \,e)$ varying 
    $y_0$, $\mu_0$ and $\theta_S$\mathversion{normal}} 
    in the ranges in eqs.~(\ref{eq:y0range}), (\ref{eq:mu0range}) and (\ref{eq:thetaSrange}), respectively. The parameters $n$, $m$ and $s$ are set to $n=34$, $m=2$ and $s=2$. 
    The light neutrino mass ordering is fixed to NO and $m_0$
    to $m_0=10^{-5} \, \mathrm{eV}$. The colour-coding of the data points is the same as in fig.~\ref{fig:Case1n26s1NOm0003}. 
    }
    \label{fig:Case3an34s2m2NOm010em5_taudecays}
\end{figure}

In fig.~\ref{fig:Case3an34s2m2NOm010em5_taudecays} results for the BRs of the four studied cLFV tau lepton decays $\tau\to\mu \, \gamma$, $\tau\to 3 \,\mu$, $\tau\to e \, \gamma$ and $\tau\to 3 \,e$ are displayed
for Case 3 a), $n=34$, $m=2$ and $s=2$, light neutrino masses with NO and $m_0=10^{-5} \, \mathrm{eV}$. The plots for IO light neutrino masses look similar, but do not reveal the feature of a (slightly) visible
    separation of the data points according to the two different viable values of the angle $\theta_N$ for the BRs of $\tau\to e \, \gamma$ and $\tau\to 3 \, e$ (for small values of $y_0$).
 Fig.~\ref{fig:Case3an34s2m2NOm010em5_taudecays} supplements the discussion in section~\ref{taucLFVs}.


\newpage

 \begin{thebibliography}{00}

 \bibitem{typeIseesaw1}
  P.~Minkowski,
Phys. Lett. B \textbf{67} (1977), 421-428.

 \bibitem{typeIseesaw2}
 T.~Yanagida, in {\it Proceedings of the Workshop on the Unified Theory and the Baryon Number in the Universe}
 (O.~Sawada and A.~Sugamoto, eds.), KEK Tsukuba, Japan, 1979, p. 95.
 
  \bibitem{typeIseesaw3}
 S.~L.~Glashow, {\it The future of elementary particle physics}, in {\it Proceedings of the 1979 Carg\`ese Summer 
 Institute on Quarks and Leptons} (M.~L\'evy, J.-L.~Basdevant, D.~Speiser, J.~Weyers, R.~Gastmans, and M.~Jacob, eds.),
 Plenum Press, New York, 1980, pp. 687-713.
  
   \bibitem{typeIseesaw4}
 M.~Gell-Mann, P.~Ramond, and R.~Slansky, {\it Complex spinors and unified theories}, in {\it Supergravity} (P.~van~Nieuwenhuizen
 and D.~Z.~Freedman, eds.), North Holland, Amsterdam, 1979, p. 315.
 
  \bibitem{typeIseesaw5}
 R.~N.~Mohapatra and G.~Senjanovic,
  Phys.\ Rev.\ Lett.\  {\bf 44} (1980) 912.
  
  \bibitem{typeIIseesaw1} 
  M.~Magg and C.~Wetterich,
Phys. Lett. B \textbf{94} (1980), 61-64.

  \bibitem{typeIIseesaw2} 
J.~Schechter and J.~W.~F.~Valle,
Phys. Rev. D \textbf{22} (1980), 2227.

  \bibitem{typeIIseesaw3} 
T.~P.~Cheng and L.~F.~Li,
Phys. Rev. D \textbf{22} (1980), 2860.

  \bibitem{typeIIseesaw4} 
G.~Lazarides, Q.~Shafi and C.~Wetterich,
Nucl. Phys. B \textbf{181} (1981), 287-300.

  \bibitem{typeIIseesaw5} 
C.~Wetterich,
Nucl. Phys. B \textbf{187} (1981), 343-375.

  \bibitem{typeIIseesaw6} 
R.~N.~Mohapatra and G.~Senjanovic,
Phys. Rev. D \textbf{23} (1981), 165.  

 \bibitem{typeIIIseesaw}
R.~Foot, H.~Lew, X.~G.~He and G.~C.~Joshi,
Z. Phys. C \textbf{44} (1989), 441.

\bibitem{Shaposhnikov:2006nn}
M.~Shaposhnikov,
Nucl. Phys. B \textbf{763} (2007), 49-59
[arXiv:hep-ph/0605047 [hep-ph]].

\bibitem{Kersten:2007vk}
J.~Kersten and A.~Y.~Smirnov,
Phys. Rev. D \textbf{76} (2007), 073005
[arXiv:0705.3221 [hep-ph]].

\bibitem{Moffat:2017feq}
K.~Moffat, S.~Pascoli and C.~Weiland,
 arXiv:1712.07611 [hep-ph].

\bibitem{Mohapatra:1986bd}
R.~N.~Mohapatra and J.~W.~F.~Valle,
Phys. Rev. D \textbf{34} (1986), 1642.

\bibitem{Mohapatra:1986aw}
R.~N.~Mohapatra,
Phys. Rev. Lett. \textbf{56} (1986), 561-563.

\bibitem{Bernabeu:1987gr}
J.~Bernabeu, A.~Santamaria, J.~Vidal, A.~Mendez and J.~W.~F.~Valle,
Phys. Lett. B \textbf{187} (1987), 303-308.

\bibitem{GonzalezGarcia:1988rw}
M.~C.~Gonzalez-Garcia and J.~W.~F.~Valle,
Phys. Lett. B \textbf{216} (1989), 360-366.

  \bibitem{GfCP}
 F.~Feruglio, C.~Hagedorn and R.~Ziegler,
  JHEP {\bf 1307} (2013) 027
  [arXiv:1211.5560 [hep-ph]]. 
  
  \bibitem{GfCPothers1}
 M.~Holthausen, M.~Lindner and M.~A.~Schmidt,
  JHEP {\bf 1304} (2013) 122
  [arXiv:1211.6953 [hep-ph]].
  
   \bibitem{GfCPothers2} 
M.-C.~Chen, M.~Fallbacher, K.~T.~Mahanthappa, M.~Ratz and A.~Trautner,
  Nucl.\ Phys.\ B {\bf 883} (2014) 267
  [arXiv:1402.0507 [hep-ph]]. 

 \bibitem{GfCPearly1}
 G.~Ecker, W.~Grimus and H.~Neufeld,
  Nucl.\ Phys.\ B {\bf 247} (1984) 70.
  
  \bibitem{GfCPearly2} 
G.~Ecker, W.~Grimus and H.~Neufeld,
  J.\ Phys.\ A {\bf 20} (1987) L807.
  
 \bibitem{GfCPearly3}  
H.~Neufeld, W.~Grimus and G.~Ecker,
  Int.\ J.\ Mod.\ Phys.\ A {\bf 3} (1988) 603.
  
  \bibitem{GfCPearly4} 
  W.~Grimus and M.~N.~Rebelo,
  Phys.\ Rept.\  {\bf 281} (1997) 239
  [arXiv:hep-ph/9506272].
  
  \bibitem{GfCPearly5} 
  P.~F.~Harrison and W.~G.~Scott,
  Phys.\ Lett.\ B {\bf 535} (2002) 163
  [arXiv:hep-ph/0203209].
  
 \bibitem{GfCPearly6}  
W.~Grimus and L.~Lavoura,
  Phys.\ Lett.\ B {\bf 579} (2004) 113
  [arXiv:hep-ph/0305309]. 
  
  \bibitem{King:2015aea}
S.~F.~King,
J. Phys. G \textbf{42} (2015), 123001
[arXiv:1510.02091 [hep-ph]].

\bibitem{Feruglio:2019ybq}
F.~Feruglio and A.~Romanino,
Rev. Mod. Phys. \textbf{93} (2021) no.1, 015007
[arXiv:1912.06028 [hep-ph]].

\bibitem{Ishimori:2010au}
H.~Ishimori, T.~Kobayashi, H.~Ohki, Y.~Shimizu, H.~Okada and M.~Tanimoto,
Prog. Theor. Phys. Suppl. \textbf{183} (2010), 1-163
[arXiv:1003.3552 [hep-th]].
  
\bibitem{Gfreviewnew}  
 T.~Kobayashi, H.~Ohki, H.~Okada, Y.~Shimizu and M.~Tanimoto,
{\it An Introduction to Non-Abelian Discrete Symmetries for Particle Physicists},
2022,
ISBN 978-3-662-64678-6, 978-3-662-64679-3.
 
\bibitem{Grimus:2011fk}
W.~Grimus and P.~O.~Ludl,
J. Phys. A \textbf{45} (2012), 233001
[arXiv:1110.6376 [hep-ph]].  
  
\bibitem{Luhn:2007uq}
C.~Luhn, S.~Nasri and P.~Ramond,
J. Math. Phys. \textbf{48} (2007), 073501
[arXiv:hep-th/0701188 [hep-th]].

\bibitem{Escobar:2008vc}
J.~A.~Escobar and C.~Luhn,
J. Math. Phys. \textbf{50} (2009), 013524
[arXiv:0809.0639 [hep-th]].  
  
 \bibitem{Hagedorn:2014wha}
C.~Hagedorn, A.~Meroni and E.~Molinaro,
Nucl. Phys. B \textbf{891} (2015), 499-557
[arXiv:1408.7118 [hep-ph]]. 
  
  \bibitem{Hagedorn:2021ldq}
C.~Hagedorn, J.~Kriewald, J.~Orloff and A.~M.~Teixeira,
Eur. Phys. J. C \textbf{82} (2022) no.3, 194
[arXiv:2107.07537 [hep-ph]].

\bibitem{DeltaCPothers1}
 G.~J.~Ding, S.~F.~King and T.~Neder,
  JHEP {\bf 1412} (2014) 007
  [arXiv:1409.8005 [hep-ph]].
  
\bibitem{DeltaCPothers2}  
G.~J.~Ding and S.~F.~King,
  Phys.\ Rev.\ D {\bf 93} (2016) 025013
  [arXiv:1510.03188 [hep-ph]].
 
\bibitem{DeltaCPothers3} 
S.~F.~King and T.~Neder,
  Phys.\ Lett.\ B {\bf 736} (2014) 308
  [arXiv:1403.1758 [hep-ph]].
  
\bibitem{DeltaCPothers4}   
 G.~J.~Ding, S.~F.~King, C.~Luhn and A.~J.~Stuart,
  JHEP {\bf 1305} (2013) 084
  [arXiv:1303.6180 [hep-ph]].
  
  \bibitem{DeltaCPothers5}   
  F.~Feruglio, C.~Hagedorn and R.~Ziegler,
  Eur.\ Phys.\ J.\ C {\bf 74} (2014) 2753
  [arXiv:1303.7178 [hep-ph]].
  
  \bibitem{DeltaCPothers6}   
  G.~J.~Ding, S.~F.~King and A.~J.~Stuart,
  JHEP {\bf 1312} (2013) 006
  [arXiv:1307.4212 [hep-ph]].
  
 \bibitem{DeltaCPothers7}    
  C.~C.~Li and G.~J.~Ding,
  Nucl.\ Phys.\ B {\bf 881} (2014) 206
  [arXiv:1312.4401 [hep-ph]].
  
\bibitem{DeltaCPothers8}     
C.~C.~Li and G.~J.~Ding,
  JHEP {\bf 1508} (2015) 017
  [arXiv:1408.0785 [hep-ph]].
  
  \bibitem{DeltaCPothers9}    
  G.~J.~Ding and Y.~L.~Zhou,
  Chin.\ Phys.\ C {\bf 39} (2015) 2,  021001
  [arXiv:1312.5222 [hep-ph]].
  
 \bibitem{DeltaCPothers10}    
 G.~J.~Ding and Y.~L.~Zhou,
  JHEP {\bf 1406} (2014) 023
  [arXiv:1404.0592 [hep-ph]].
  
 \bibitem{DeltaCPothers11}    
   G.~J.~Ding and S.~F.~King,
  Phys.\ Rev.\ D {\bf 89} (2014) 9,  093020
  [arXiv:1403.5846 [hep-ph]]. 

\bibitem{DiMeglio:2024gve}
F.~P.~Di Meglio and C.~Hagedorn,
Nucl. Phys. B \textbf{1018} (2025), 117020
[arXiv:2407.19734 [hep-ph]].

\bibitem{Drewes:2022kap}
M.~Drewes, Y.~Georis, C.~Hagedorn and J.~Klari\'c,
JHEP \textbf{12} (2022), 044
[arXiv:2203.08538 [hep-ph]].

 \bibitem{Hettmansperger:2011bt}
H.~Hettmansperger, M.~Lindner and W.~Rodejohann,
JHEP \textbf{04} (2011), 123
[arXiv:1102.3432 [hep-ph]].
  
 \bibitem{Esteban:2024eli}    
 I.~Esteban, M.~C.~Gonzalez-Garcia, M.~Maltoni, I.~Martinez-Soler, J.~P.~Pinheiro and T.~Schwetz,
JHEP \textbf{12} (2024), 216
[arXiv:2410.05380 [hep-ph]]. 

  \bibitem{Planck:2018vyg}
N.~Aghanim \textit{et al.} [Planck],
Astron. Astrophys. \textbf{641} (2020), A6
[erratum: Astron. Astrophys. \textbf{652} (2021), C4]
[arXiv:1807.06209 [astro-ph.CO]].  
  
\bibitem{Alonso:2012ji}
R.~Alonso, M.~Dhen, M.~B.~Gavela and T.~Hambye,
JHEP \textbf{01} (2013), 118
[arXiv:1209.2679 [hep-ph]]. 
 
 \bibitem{Ilakovac:1994kj}
A.~Ilakovac and A.~Pilaftsis,
Nucl. Phys. B \textbf{437} (1995), 491
[arXiv:hep-ph/9403398].  

\bibitem{Kitano:2002mt}
R.~Kitano, M.~Koike and Y.~Okada,
Phys. Rev. D \textbf{66} (2002), 096002
[erratum: Phys. Rev. D \textbf{76} (2007), 059902]
[arXiv:hep-ph/0203110 [hep-ph]].
  
\bibitem{delAguila:2008cj}
F.~del Aguila and J.~A.~Aguilar-Saavedra,
Nucl. Phys. B \textbf{813} (2009), 22-90
[arXiv:0808.2468 [hep-ph]].

\bibitem{delAguila:2008hw}
F.~del Aguila and J.~A.~Aguilar-Saavedra,
Phys. Lett. B \textbf{672} (2009), 158-165
[arXiv:0809.2096 [hep-ph]].

\bibitem{Chen:2011hc}
C.~Y.~Chen and P.~S.~B.~Dev,
Phys. Rev. D \textbf{85} (2012), 093018
[arXiv:1112.6419 [hep-ph]].

\bibitem{Das:2012ze}
A.~Das and N.~Okada,
Phys. Rev. D \textbf{88} (2013), 113001
[arXiv:1207.3734 [hep-ph]].

\bibitem{Abada:2014vea}
A.~Abada and M.~Lucente,
Nucl. Phys. B \textbf{885} (2014), 651-678
[arXiv:1401.1507 [hep-ph]].

\bibitem{Arganda:2014dta}
E.~Arganda, M.~J.~Herrero, X.~Marcano and C.~Weiland,
Phys. Rev. D \textbf{91} (2015) no.1, 015001
[arXiv:1405.4300 [hep-ph]].

\bibitem{Abada:2014nwa}
A.~Abada, V.~De Romeri and A.~M.~Teixeira,
JHEP \textbf{09} (2014), 074
[arXiv:1406.6978 [hep-ph]].

\bibitem{Abada:2014kba}
A.~Abada, M.~E.~Krauss, W.~Porod, F.~Staub, A.~Vicente and C.~Weiland,
JHEP \textbf{11} (2014), 048
[arXiv:1408.0138 [hep-ph]].

\bibitem{Abada:2014cca}
A.~Abada, V.~De Romeri, S.~Monteil, J.~Orloff and A.~M.~Teixeira,
JHEP \textbf{04} (2015), 051
[arXiv:1412.6322 [hep-ph]].

\bibitem{Arganda:2015naa}
E.~Arganda, M.~J.~Herrero, X.~Marcano and C.~Weiland,
Phys. Rev. D \textbf{93} (2016) no.5, 055010
[arXiv:1508.04623 [hep-ph]].
 
 \bibitem{Abada:2015oba}
A.~Abada, V.~De Romeri and A.~M.~Teixeira,
JHEP \textbf{02} (2016), 083
[arXiv:1510.06657 [hep-ph]].
 
 \bibitem{DeRomeri:2016gum}
V.~De Romeri, M.~J.~Herrero, X.~Marcano and F.~Scarcella,
Phys. Rev. D \textbf{95} (2017) no.7, 075028
[arXiv:1607.05257 [hep-ph]].
 
 \bibitem{Antusch:2016ejd}
S.~Antusch, E.~Cazzato and O.~Fischer,
Int. J. Mod. Phys. A \textbf{32} (2017) no.14, 1750078
[arXiv:1612.02728 [hep-ph]].

\bibitem{Crivellin:2022cve}
A.~Crivellin, F.~Kirk and C.~A.~Manzari,
JHEP \textbf{12} (2022), 031
[arXiv:2208.00020 [hep-ph]].

\bibitem{Abada:2024hpb}
A.~Abada and T.~Toma,
JHEP \textbf{08} (2024), 128
[arXiv:2405.01648 [hep-ph]].

\bibitem{Blennow:2023mqx}
M.~Blennow, E.~Fern\'andez-Mart\'\i{}nez, J.~Hern\'andez-Garc\'\i{}a, J.~L\'opez-Pav\'on, X.~Marcano and D.~Naredo-Tuero,
JHEP \textbf{08} (2023), 030
[arXiv:2306.01040 [hep-ph]].

\bibitem{MEGII:2025gzr}
K.~Afanaciev \textit{et al.} [MEG II],
Eur. Phys. J. C \textbf{85} (2025) no.10, 1177
[erratum: Eur. Phys. J. C \textbf{85} (2025) no.11, 1317]
[arXiv:2504.15711 [hep-ex]].

\bibitem{SINDRUM:1987nra}
U.~Bellgardt \textit{et al.} [SINDRUM],
Nucl. Phys. B \textbf{299} (1988), 1-6.

 \bibitem{SINDRUMII:2006dvw}
W.~H.~Bertl \textit{et al.} [SINDRUM II],
Eur. Phys. J. C \textbf{47} (2006), 337-346.
 
 \bibitem{SINDRUMIITi}
 P. Wintz, 
 {\it Prepared for 29th International Conference on High-Energy Physics (ICHEP 98), 
 Vancouver, Canada, 23-29 Jul 1998.}

\bibitem{MEGII:2021fah}
A.~M.~Baldini \textit{et al.} [MEG II],
Symmetry \textbf{13} (2021) no.9, 1591
[arXiv:2107.10767 [hep-ex]].
 
 \bibitem{Blondel:2013ia}
A.~Blondel, A.~Bravar, M.~Pohl, S.~Bachmann, N.~Berger, M.~Kiehn, A.~Schoning, D.~Wiedner, B.~Windelband, P.~Eckert \textit{et al.},
arXiv:1301.6113 [physics.ins-det].

\bibitem{COMET:2018auw}
R.~Abramishvili \textit{et al.} [COMET],
PTEP \textbf{2020} (2020) no.3, 033C01
[arXiv:1812.09018 [physics.ins-det]].

\bibitem{Jansen:2023ojv}
A.~Jansen [COMET],
EPJ Web Conf. \textbf{282} (2023), 01014.
 
 \bibitem{Mu2e:2014fns}
L.~Bartoszek \textit{et al.} [Mu2e],
arXiv:1501.05241 [physics.ins-det].

\bibitem{Artuso:2022ouk}
M.~Artuso, R.~H.~Bernstein, A.~A.~Petrov, T.~Blum, A.~Di Canto, S.~Davidson, B.~Echenard, S.~Gori, E.~Goudzovski, R.~F.~Lebed \textit{et al.},
arXiv:2210.04765 [hep-ex].

 \bibitem{Belle:2021ysv}
A.~Abdesselam \textit{et al.} [Belle],
JHEP \textbf{10} (2021), 19
[arXiv:2103.12994 [hep-ex]].

\bibitem{ParticleDataGroup:2020ssz}
P.~A.~Zyla \textit{et al.} [Particle Data Group],
PTEP \textbf{2020} (2020) no.8, 083C01.

 \bibitem{Belle-II:2024sce}
I.~Adachi \textit{et al.} [Belle II],
JHEP \textbf{09} (2024), 062
[arXiv:2405.07386 [hep-ex]].

\bibitem{LHCb:2026eod}
R.~Aaij \textit{et al.} [LHCb],
arXiv:2601.20785 [hep-ex].

\bibitem{Banerjee:2022xuw}
S.~Banerjee, V.~Cirigliano, M.~Dam, A.~Deshpande, L.~Fiorini, K.~Fuyuto, C.~Gal, T.~Husek, E.~Mereghetti, K.~Mons\'alvez-Pozo \textit{et al.},
arXiv:2203.14919 [hep-ph].

\end{thebibliography}
\end{document}